\documentclass{aa}  

\usepackage{graphicx}
\usepackage{txfonts}
\usepackage{natbib}
\bibpunct{(}{)}{;}{a}{}{,} % to follow the A&A style

\begin{document}

   \title{\textit{Chandra}\thanks{The \textit{Chandra} data described in this paper
have been obtained in the open time project with
ObsID 16648 (PI: T.~Preibisch) ivo://ADS/Sa.CXO\#obs/16648.} 
X-ray observation of the young stellar cluster NGC~3293 in the 
Carina Nebula Complex\thanks{Tables 1, 2, and 3 are only available in electronic form
at the CDS via anonymous ftp to cdsarc.u-strasbg.fr (130.79.128.5)
or via http://cdsweb.u-strasbg.fr/cgi-bin/qcat?J/A+A/}}

   \author{T.~Preibisch\inst{1}
          \and S.~Flaischlen\inst{1}
          \and B.~Gaczkowski\inst{1}
          \and L.~Townsley\inst{2}
          \and P.~Broos\inst{2} 
          }

   \institute{Universit\"ats-Sternwarte M\"unchen, 
              Ludwig-Maximilians-Universit\"at,
              Scheinerstr.~1, 81679 M\"unchen, Germany\\
              \email{preibisch@usm.uni-muenchen.de}
\and
Department of Astronomy \& Astrophysics,
             Pennsylvania State University, University Park PA 16802, USA
             }

\titlerunning{\textit{Chandra} X-ray observation of NGC~3293}
\authorrunning{Preibisch et al.}

   \date{Received March 27, 2017; accepted July 11, 2017}

% \abstract{}{}{}{}{} 
% 5 {} token are mandatory
 
  \abstract
   {NGC~3293 is a young stellar cluster at the northwestern
    periphery of the Carina Nebula Complex that has remained poorly explored until now.
   }
   {
    We  characterize the stellar population of NGC~3293
    in order to evaluate key parameters of the cluster population such as  the age and the mass function,
  and to test claims of an abnormal IMF and a deficit of $M \le 2.5\,M_{\odot}$ stars.
   }
   {We performed a deep (70 ksec) X-ray observation of NGC~3293 with
   \textit{Chandra}
 and detected 1026 individual X-ray point sources. 
These X-ray data directly probe the low-mass ($M \leq 2\,M_\odot$) 
stellar population
by means of the strong X-ray emission of young low-mass stars.
We identify counterparts for 74\% of the X-ray sources in our deep near-infrared images.
   }
   {Our data clearly show that NGC~3293 hosts a large population
of $\approx$~solar-mass stars, refuting claims of a lack of $M \le 2.5\,M_\odot$ stars.
The analysis of the color magnitude diagram suggests an age of $\sim 8-10$~Myr
for the low-mass population of the cluster. 
There are at least 511 X-ray detected stars with color magnitude positions that are
consistent with young stellar members within 7 arcmin
of the cluster center.
The number ratio of X-ray detected stars in the $[1-2]\,M_\odot$ range
versus the $M \ge 5\,M_\odot$ stars (known from optical spectroscopy)
is  consistent with the expectation from a normal
field initial mass function. Most of the early B-type stars and $\approx 20\%$ of the
later B-type stars are detected as X-ray sources.
   }
   {
Our data shows that NGC~3293 is one of the most populous stellar clusters
in the entire Carina Nebula Complex (very similar to 
Tr~16 and Tr~15;  only  Tr~14 is more populous).
The cluster  probably harbored several O-type stars, whose
supernova explosions  may have had an important impact on the early evolution
of the Carina Nebula Complex.
}


   \maketitle

%%%%%%%%%%%%%%%%%%%%%%%%%%%%%%%%%%%%%%%%%%%%%%%%%%%%%%%%%%%%%%%%

\section{Introduction}

\begin{figure*}
\parbox[t]{18.0cm}{
\parbox[c]{9.1cm}{\includegraphics[width=9.1cm]{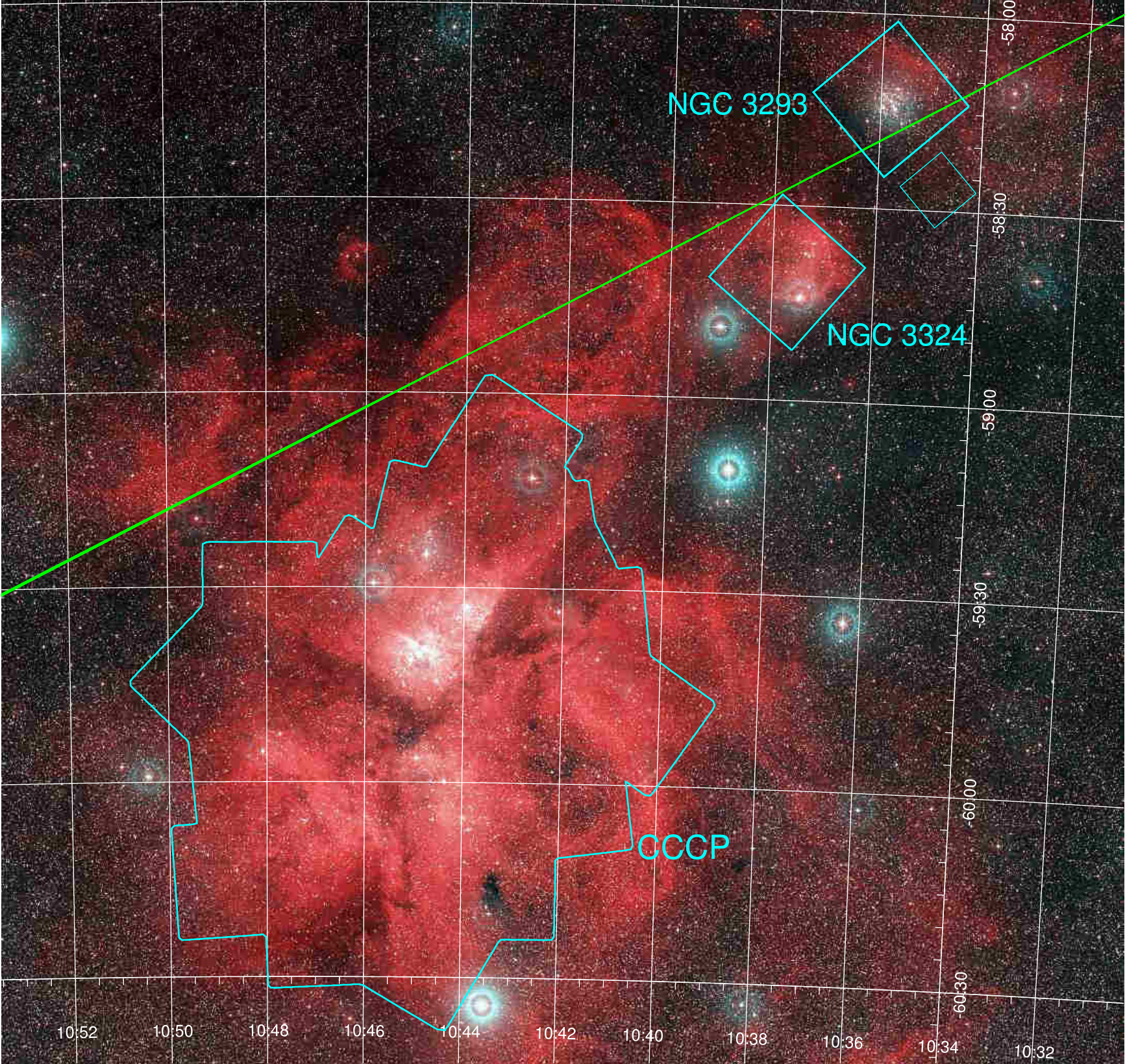}}\hspace{3mm}
\parbox[c]{8.5cm}{\includegraphics[width=8.5cm]{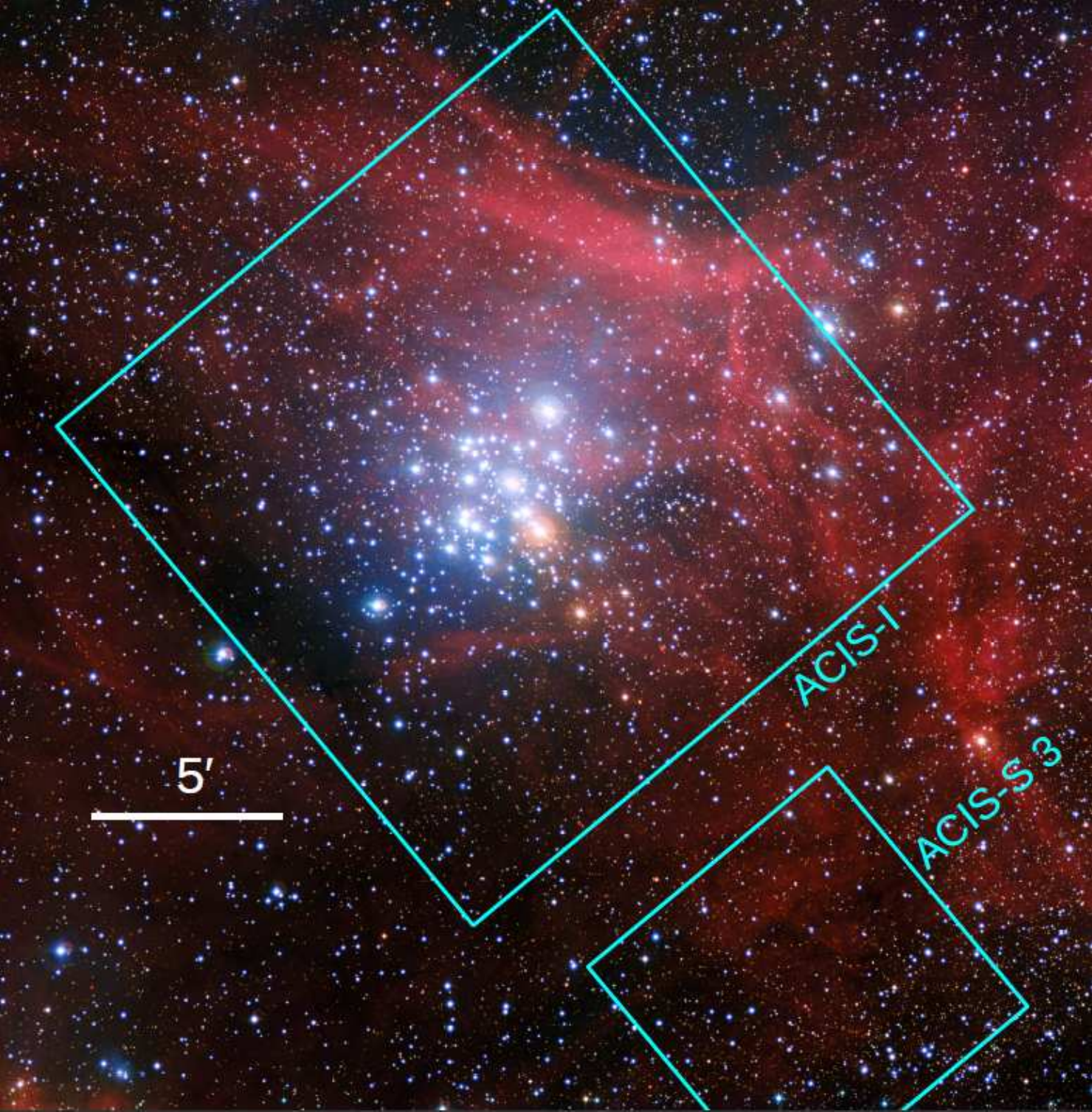}}}
\caption{Left:
Optical image of the Carina Nebula Complex
(from: {\tt www.eso.org/public/images/eso0905b/}; image credit: ESO/Digitized Sky Survey 2, 
Davide De Martin). North is up and east to the left.
The region  observed in the context of the \textit{Chandra} Carina Complex
Project (CCCP) and the \textit{Chandra} pointings of NGC~3324/Gum~31
and NGC~3293 are marked by the cyan outlines and labeled.\newline
Right: Optical image of the cluster NGC~3293 (from: {\tt www.eso.org/public/images/eso1422a/}; 
image credit: ESO/G. Beccari),
composed from images taken with the MPG/ESO 2.2m telescope
in the $B$-band (blue color channel), $V$-band (green color channel),
$I$-band (orange color channel), and an H$\alpha$ filter (red color channel).
North is up and east to the left.
The central cyan rectangle marks the area covered by our \textit{Chandra}
observation with the ACIS-I array, the lower cyan rectangle (extending beyond
the edge of the image) marks the area covered by the single active ACIS-S
detector. 
\label{CNC-3293.fig}}
\end{figure*}

The Carina Nebula Complex \citep[CNC; see][for a review]{SB08}
is one of the most massive and active star forming regions in our Galaxy.
At a moderate and well-known distance of 2.3~kpc \citep{Smith06}, 
the spatial extent of the nebulosity of about 100~pc corresponds to
several degrees on the sky. An optical image of the CNC is shown in 
Fig.~\ref{CNC-3293.fig}.
The clouds contain a total gas and dust mass of about $10^6\,M_\odot$
\citep{Preibisch12} and harbor more than
100\,000 young stars 
\citep{CCCP-Clusters,Povich11,HAWKI-survey,VISTA2}.
Most of these young stars are located in one of several
clusters, including the extensively studied clusters
Tr~14, 15, and 16 in the central regions
of the Carina Nebula, and the clusters NGC~3324 and NGC~3293 in the
northern part of the CNC. 

Over the last years, several sensitive surveys
of large parts of the CNC have been performed at optical and infrared wavelengths 
\citep[e.g.,][]{Smith10a,Smith10b,CNC-Laboca,HAWKI-survey,Preibisch12,VISTA1}.
However, due to the location of the 
CNC very close to the Galactic plane, all optical and infrared
images are completely
dominated by unrelated field stars in the galactic background, which presents
a major obstacle in the identification of the young stellar population
in the complex.
X-ray observations provide a
very good way to solve this problem because strong X-ray emission is
a very good tracer of stellar youth and allows very  young stars to be efficiently distinguished from the numerous (and much older) field
stars in the galactic background \citep[see][]{Feigelson07}.
A major milestone in the exploration of the CNC was therefore the
deep X-ray imaging survey of the
{\it Chandra} Carina Complex Project \citep[CCCP; see][for an overview]{CCCP-intro},
which mapped the central 1.4 square degrees 
(i.e.,~roughly half of the total spatial extent of the CNC)
with a mosaic of 22 individual  pointings with the
Imaging Array of the \textit{Chandra}
Advanced CCD Imaging Spectrometer \citep[ACIS-I; see][]{Garmire03}.
The \textit{Chandra} data revealed
14\,368 individual X-ray sources, 10\,714 of which
are most likely young stars in the Carina Nebula \citep{CCCP-classification}
with masses down to  $\sim 0.5\,M_\odot$.
The combination of these CCCP X-ray data with a deep near-infrared (NIR) 
survey \citep{HAWKI-survey} 
provides information about the properties of the stellar populations
in the central parts of the complex \citep{CCCP-HAWKI},
including the individual stellar
clusters Tr~16 and Tr~15 \citep{CCCP-Tr16,CCCP-Tr15}.

The northern parts of the cloud complex, including the
young stellar clusters NGC~3324 and NGC~3293,
were much less well explored 
because this area was not included in the
CCCP and, until recently,  no deep infrared survey was available.
The new very wide-field (6.7 square degrees) NIR survey with the
ESO 4m VISTA telescope \citep[see][]{VISTA1} finally solved this problem;
these data are sensitive enough to detect the full stellar population of young stars
down to $\sim 0.1\,M_\odot$ over the entire area of the CNC,
including the
less  studied northern parts.

As a first step in the X-ray exploration of the young stellar populations in these
northern parts of the CNC (outside the field covered by the CCCP),
we  recently obtained a \textit{Chandra} ACIS-I pointing of the
stellar cluster NGC~3324 in the prominent HII region Gum~31 \citep{Preibisch14},
which led to the detection of 679 new X-ray sources.

The only cluster in the CNC that had not yet been observed in X-rays 
was NGC~3293 (also known as the ``Gem Cluster''). Although this  cluster  
is a very prominent celestial object (see Fig.~\ref{CNC-3293.fig}),
it was often neglected in studies of the CNC
because of its angular separation of $\sim 1.5\degr$ (70~pc)
from the center of the Carina Nebula.
Recent distance estimates for NGC~3293 (based on optical photometry)
of
$\approx 2327$~pc \citep{Dias02} and
$\approx 2471$~pc \citep{Kharchenko05} are very
consistent (within their uncertainties) with the
distance of the Carina Nebula (2.3~kpc). 
The extinction for most stars in the cluster center is quite low
($A_V \la 1$~mag), but increases up to several magnitudes for stars
more than a few arcmin away from the cluster center, due to 
the interstellar clouds around the cluster (see Appendix 
\ref{clouds-extinction.sec}).

NGC~3293 contains an impressive collection of bright
stars, including 48 early B-type stars 
\citep{Evans05} and several supergiants,
e.g.,~HD~91969 (B0Ib) and V361~Car (M1.5Iab).
The presence of evolved massive stars, main-sequence stars up
to spectral type B0.5~V, and 
comparisons of optical color magnitude diagrams with stellar 
evolution models  suggested
an age of $\approx 8-10$~Myr for the high-mass population in 
this cluster \citep{Slawson07,Delgado11}.

Another reason that makes NGC~3293 particularly interesting
are claims for an abnormal stellar
initial mass function (IMF) and an apparent severe lack
of low-mass stars. 
In their analysis of optical photometry,
\citet{Slawson07} came to the conclusion
that the cluster's IMF shows
a sharp turnover at masses below $2.4\,M_\odot$.
They argue that NGC~3293 is the cluster with the most
convincing evidence for a strong deficit of low-mass stars.
 \citet{Delgado11} also
arrived at a similar conclusion from their optical photometry study,  which suggested
that the mass function is
considerably flatter than the Salpeter slope below $2.5\,M_\odot$.

These results, however,  as well as all the other studies of the low-mass
population of NGC~3293 performed so far, were only based
on photometric observations. In the analysis of color magnitude diagrams of stellar clusters,
the statistical background subtraction can introduce a very serious source
of uncertainties, in particular at locations close to the galactic plane.
Much more reliable information about the cluster properties 
and mass function can be
obtained if the low-mass stellar members of the cluster can be individually
identified.

The main obstacle for a reliable identification of the 
low-mass stellar population
of NGC~3293 is the confusion by the
very strong field star contamination resulting from the cluster's location
very close to the galactic plane: the cluster center has a 
galactic latitude of $b = +0.08^\circ$.
All optical and infrared images are therefore strongly contaminated
by unrelated field stars.
 The  approach most often used  to identify young stars is by
the infrared excess emission from circumstellar disks, which
 is not feasible here.  At an age of $\sim 8$~Myr, most stars should have already
lost their circumstellar accretion disks and thus do not exhibit infrared
excesses \citep[e.g.,][]{Fedele10}. 
It is thus essentially impossible to identify and distinguish a population
of $\sim 8-10$~Myr old low-mass stars from
unrelated field stars with optical or infrared photometry alone.

\textit{Chandra}, however, can  solve this problem because 
the strongly enhanced
X-ray emission of young stars \citep{Feigelson07,Preibisch_coup_orig}
provides an extremely useful
way to discriminate  between
young pre-main-sequence stars and the much older field stars.
The median X-ray luminosity of $\sim 10$~Myr old
solar-mass stars is nearly 1000 times higher than
for solar-mass field stars (see Preibisch \& Feigelson 2005),
and makes these young stars relatively easy targets to  detect
for \textit{Chandra} even at a distance of 2.3~kpc.
A \textit{Chandra} observation of NGC~3293
can uncover an X-ray luminosity-limited (i.e.,~approximately
mass-limited) sample of the low-mass stellar population,
provide a comprehensive census of the cluster members,
and immediately answer the open questions about the size of the
low-mass stellar population (and thus the total cluster mass).

%%%%%%%%%%%%%%%%%%%%%%%%%%%%%%%%%%%%%%%%%%%%%%%%%%%%%%%%%%%%%%%%%%%%%%%%%%%%%%%%%%%%%%%%%%%

\begin{figure*} \centering
\includegraphics[width=16.0cm]{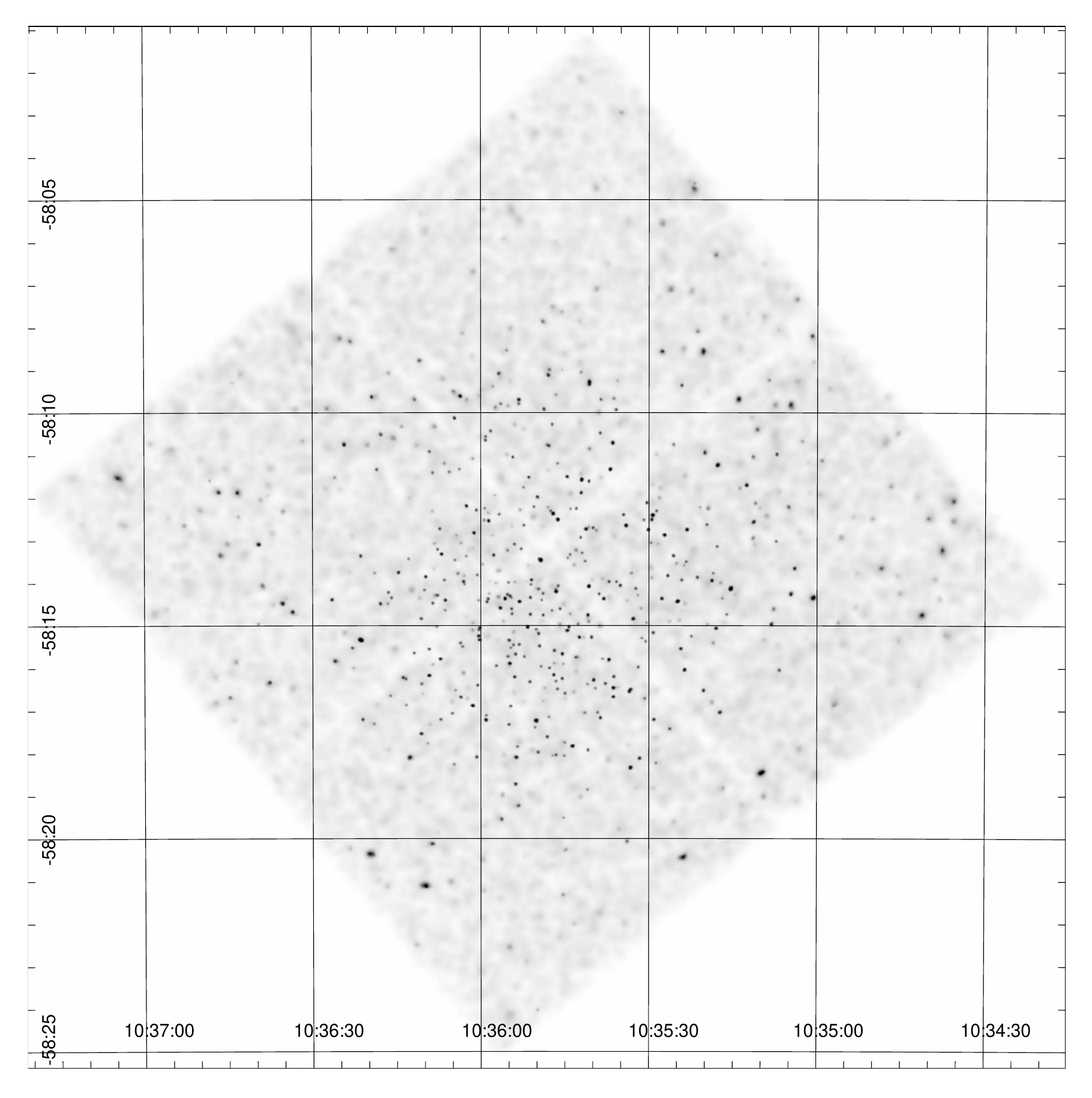}
\caption{Negative grayscale representation of the \textit{Chandra} ACIS-I 
image for the total band ($0.5-8~$keV). 
The image was smoothed with the CIAO tool \texttt{csmooth} and is 
displayed with an asinh intensity scale. North is up and east to the left.
\label{chandra-image.fig}}
\end{figure*}

\begin{figure*} \centering
\includegraphics[width=15.0cm]{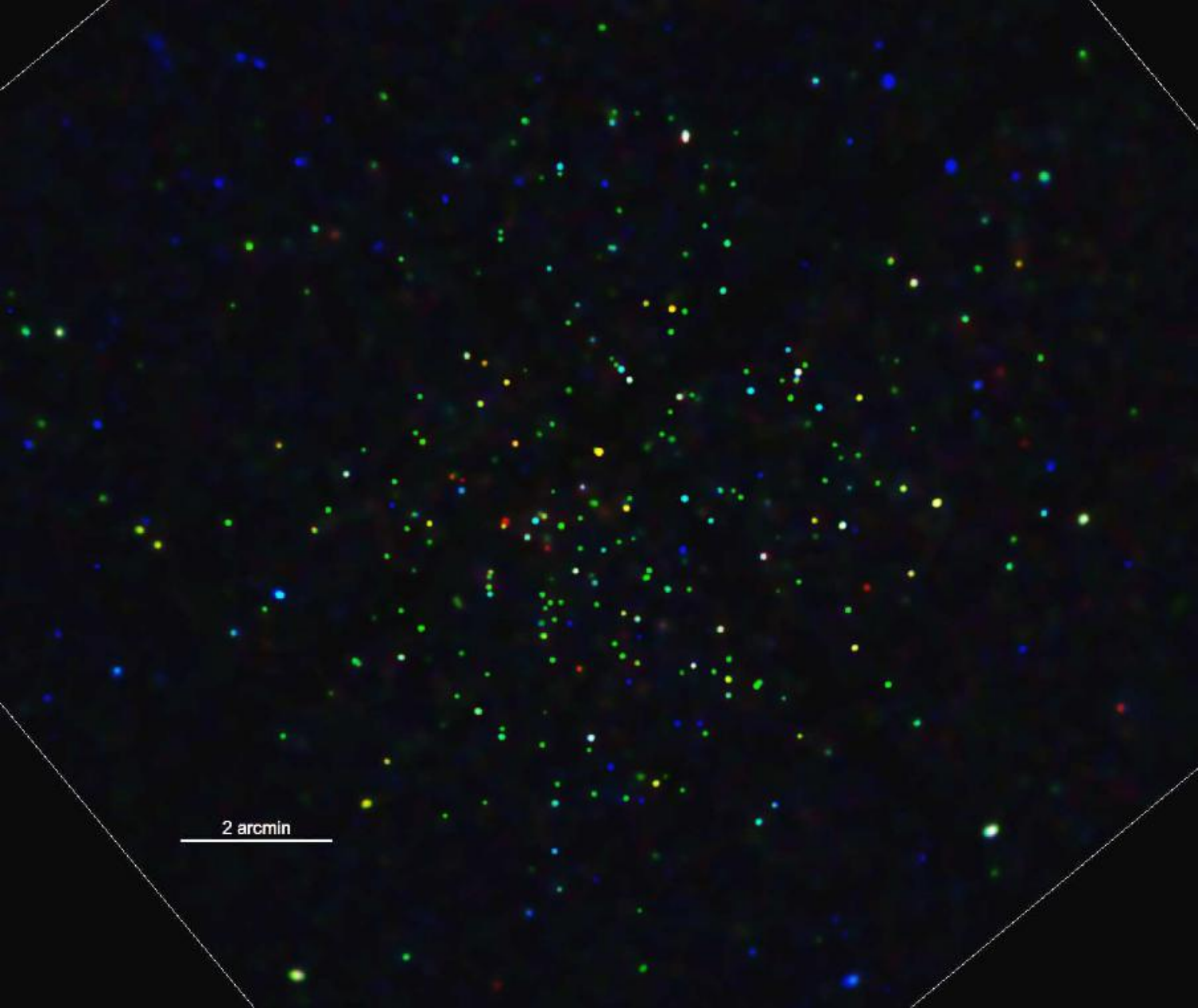}
\caption{RGB Color composite of the central part of the
\textit{Chandra} ACIS-I image in selected
energy ranges (red: $[0.5 - 1]$ keV, green: $[1 - 2]$~keV, 
blue: $[2 - 8]$~keV). The image was smoothed with the CIAO tool \texttt{csmooth} 
and is
displayed with an asinh intensity scale. North is up and east to the left.
\label{chandra-rgb.fig}}
\end{figure*}

\section{\textit{Chandra} X-ray observation and data analysis}

We  used the \textit{Chandra} X-ray observatory \citep{Weisskopf02} 
 to perform a deep pointing of the cluster NGC~3293 with the
Imaging Array of the \textit{Chandra}
Advanced CCD Imaging Spectrometer \citep[ACIS-I; see][]{Garmire03}.
The observation was performed as an open time project 
with 
ObsID 16648 (PI: T.~Preibisch)
during \textit{Chandra} Observing Cycle~15 in 
October 2015 (start date: 2015-10-07T10:14:23, 
end date: 2015-10-08T06:43:28). 
The imaging array
ACIS-I provides a field of view  of $17' \times 17'$ on the sky
(which corresponds to  $11.3 \times 11.3$~pc at the distance of
2.3~kpc),  and has a pixel size of $0.492''$.
The point spread function of the X-ray telescope
has a FWHM of $0.5''$ on-axis, but
increases towards the edge of the detector.
As part of our data analysis, the 
\textit{Chandra} data were astrometrically registered to
2MASS to improve the absolute astrometry.
The positions of bright X-ray sources can thus usually be determined
with subarcsecond precision.

The observation was performed in the standard
``Timed Event, Very Faint'' mode with $5 \times 5$ pixel event islands, and the
total effective exposure time (``livetime'') of the observation
was 70\,870~s (19.68 hours).

In addition to ACIS-I, one CCD detector (CCD 7 = S3) 
of the spectroscopic array ACIS-S was also operational during our pointing.
It covers a  $8.3' \times 8.3'$ area on the sky  southwest
 of the cluster center.
While the ACIS-I chips are front-illuminated, the S3 chip is back-illuminated,
and thus its response extends to energies below that accessible by the FI chips.
This causes a substantially higher level of background in the S3 chip.
Furthermore, the  point spread function is 
seriously degraded at the rather large off-axis angles of the S3 chip. 
These two effects lead to a considerably higher detection limit for
point sources in the area covered by the
S3 chip compared to the region covered by the ACIS-I array.
Nevertheless, the S3 data were included in our data analysis 
and source detection, and contributed four point sources to the total
source list.

The aimpoint of the observation was 
  $\alpha ({\rm J2000}) = 10^{\rm h}\,35^{\rm m}\,50.07^{\rm s}$,
$\delta ({\rm J2000}) = -58\degr\,14'\,00''$, which is
close to the optical center of the cluster (see Fig.~1).
The pointing roll angle
(i.e.,~the orientation of the detector with respect to the celestial 
north direction) was $140.19\degr$.

The resulting X-ray image is shown in Fig.~\ref{chandra-image.fig}.
Clearly hundreds of X-ray sources are detected, with the greatest concentration 
at the cluster center.  The gaps between the four ACIS-I CCDs are discernible 
as a light cross bisecting the image in two directions; the effective exposure time 
is reduced in these gaps. 
Figure \ref{chandra-rgb.fig} shows a RGB color composite version of the
\textit{Chandra} image with color-coding corresponding to photon energy.
X-ray sources can appear harder (bluer) because of their intrinsic spectral shape 
or because they suffer high absorption.

At the distance of 2.3~kpc, the expected ACIS point source sensitivity limit
for a three-count detection on-axis in a 71~ks observation is
$L_{\rm X,min} \simeq  10^{29.7}$~erg~s$^{-1}$ in the $[0.5 - 8]$~keV band,
assuming an extinction of 
$A_V \approx 1$~mag ($N_{\rm H} \approx 2 \times 10^{21}\,{\rm cm}^{-2}$)
as typical for the stars in the central region of NGC~3293,
and a thermal plasma
with $kT = 1$~keV \citep[which is a typical value for young stars; see, e.g.,][]{Preibisch_coup_orig}.
Using the  empirical relation between X-ray luminosity and stellar mass
and the temporal evolution of X-ray luminosity
from the 
sample of young stars in the Orion Nebula Cluster,
which was very well studied in the \textit{Chandra} Orion 
Ultradeep Project \citep{Preibisch_coup_orig,PF05},
we can expect to detect $\approx 90\%$ of the $\approx$ solar-mass stars  
in the central region of the young cluster NGC~3293.

\subsection{X-ray point-source detection and extraction with \textit{ACIS Extract} \label{ae_analysis.sec}}

Source detection and analysis followed the steps described 
in \citet{Broos10,AE2012} (hereafter  B10); 
we  provide a brief summary of this procedure here,
but interested readers are encouraged to review the details in B10, 
especially the caveats and warnings regarding the detection of faint X-ray sources.
\textit{Chandra}-ACIS event data were calibrated and cleaned as described 
in Section 3 of B10. 
Candidate point sources were then identified in the pointing using 
Lucy-Richardson image reconstructions  \citep{Lucy74} of small overlapping 
images that ``tile'' the field in three energy bands (0.5--7, 0.5--2, 
2--7 keV).
These candidate sources were then extracted using the most recent and
improved version of the
{\em ACIS Extract} (hereafter AE) software 
package\footnote{The {\em ACIS Extract} software package and User's Guide are available 
at {\tt http://www.astro.psu.edu/xray/acis/acis\_analysis.html}.} \citep{AE2012},
which also uses the CIAO and MARX software \citep{Fruscione06,Davis12}.
For each source, AE calculated the probabilities that the counts extracted in each of 
three energy bands arose solely from the local background.
When all three probabilities were greater than 0.01 or when less than three X-ray counts 
were extracted, the candidate source was judged to be not significant, 
and removed from the list.
The positions of surviving source candidates were updated with AE estimates, and the 
reduced list of candidates was re-extracted.
This cycle of extraction, pruning, and position estimation was repeated until no candidates 
were found to be insignificant.
Our final X-ray catalog contains 1026 individual point sources, 
1022 of which are located in the field of view of the ACIS-I array,
and 4 in the field covered by the single active ACIS-S chip.

The final list of X-ray sources with their properties
is given in the electronic Table 1
(available at the CDS). Sources are
identified by their sequence number (Col.~1)
or their IAU designation (Col.~2).
Following the rules for the
designation of sources found with the \textit{Chandra} X-ray Observatory,
we have registered the acronym
\textit{CXONGC3293} as the prefix for the IAU designation (Col.~2).

The choice of the source significance limit (denoted  ``${\rm ProbNoSrc}_{\rm min}$'' 
in the text below
and listed as $P_B$ in Table 1) for the definition of the
final sample of X-ray sources is always a compromise.
A strict limit will miss true X-ray sources, whereas a
more generous limit will enhance the risk of including spurious sources
into the sample. The limiting value we use here (${\rm ProbNoSrc}_{\rm min} = 0.01$) tends to
be on the generous side and allows the detection of 
sources with just $\sim 3$ counts. We note that the same value 
has been used in numerous similar \textit{Chandra} studies of 
young stellar clusters \citep[e.g.,][]{Feigelson13}, and in particular also in the
CCCP \citep[see][for a detailed discussion of the threshold value]{CCCP-catalog}.

In order to keep track of the possible effects of potentially spurious sources
on the analysis of the cluster properties in Sect.~4,
we defined two subsamples of our source catalog.
The first group comprises the very significant sources, defined as 
having a low probability of being a spurious source,
${\rm ProbNoSrc}_{\rm min}$ in Table 1 less than 0.003.
This group contains 849 of the 1022 ACIS-I sources, which we  refer to as
``highly reliable'' X-ray sources in the following text.
The second group comprises the complementary set of
173  less significant ACIS-I sources 
($0.003 < {\rm ProbNoSrc}_{\rm min}< 0.01$),
which we  refer to as ``less reliable'' X-ray sources in the following text.
In the analysis in the following sections we  check how the results
change when we include or exclude the subsample of the
less reliable X-ray sources.
As shown there, we find that our derived quantitative 
results about the cluster properties are not significantly dependent
on whether the less reliable X-ray sources are included or excluded.

\subsection{X-ray variability}

As part of the AE procedure, the time variability of each X-ray source
 is investigated 
by comparing the arrival times of the individual
source photons in each extraction region to a model assuming
temporal uniform count rates.
The statistical significance for variability 
is then computed with a one-sided Kolmogorov-Smirnov
statistic (Col.~15 of Table~1). In our sample,
 29 sources show significant X-ray variability (probability of being constant
$P_{\rm KS} <0.005$) and an additional 71 sources
 are classified as possibly variable 
($0.005 <  P_{\rm KS} < 0.05$).

The light curves of the variable X-ray sources show
a variety of temporal behaviors.
Some sources show
flare-like variability, i.e.,~a fast increase in the count rate followed
by a slow exponential decay,  which is typical of
solar-like magnetic reconnection flares \citep[see, e.g.,][]{Wolk05}.
Other variable sources show
more slowly increasing or decreasing count rates, as is  often found
for young stellar objects
\citep[see, e.g.,][]{Stassun06}.

\subsection{X-ray spectral analysis for bright sources \label{spectra.sec}}

Modeling the observed X-ray spectrum provides the best way 
to estimate the intrinsic X-ray luminosity of a source,
but it requires a sufficiently large number of source counts.
We restricted the  spectral fitting analysis to 
57 sources  detected with  $S/N > 4$.

XSPEC version 12.9.0i was used for this analysis.
The background-subtracted spectra were 
grouped into bins of constant signal-to-noise ratio,
and the fits were then performed using the
$\chi^2$ statistic.
Considering the relatively small number of counts in our spectra, it is
advisable to keep the number of free parameters as low as possible.
We therefore only used one-temperature models\footnote{We tried
spectral models with more than one temperature component
for the brightest sources, but found that such more complicated models
were not a statistical improvement on the
one-temperature fits.},
where the thermal plasma was described with the {\it VAPEC} model,
and the 
{\it TBABS} model described the effect of
extinction by interstellar and/or circumstellar material (as
parameterized by the hydrogen column density $N_{\rm H}$).
The elemental abundances for the plasma model were fixed at the
values\footnote{The adopted abundances, relative to the solar photospheric
abundances given by \citet{ag89}, are:
C = 0.45, N = 0.788, O = 0.426, Ne = 0.832, Mg = 0.263, Al = 0.5, 
Si = 0.309, S = 0.417, Ar = 0.55, Ca = 0.195, Fe = 0.195, Ni = 0.195.}
that were found by \citet{Guedel07} to be typical
for young stellar objects.

In our fitting, the $N_{\rm H}$ parameter was allowed to vary between
$10^{19}\,{\rm cm}^{-2}$  and $10^{23}\,{\rm cm}^{-2}$ and 
the plasma temperature $kT$ between 0.1  and 10~keV.
Since the parameter space of these fits can have a complicated
topology rather than a clearly defined, unique minimum, 
we performed at least six fits for each source
with different values of the starting parameters.
The grid of starting parameter values is spanned by the
vectors
$N_{\rm H} = [10^{21}, 10^{22}]\,{\rm cm}^{-2}$ and $kT = [0.3, 1.0, 3.0]$~keV.
From the resulting fits, the model with the overall 
highest value of the null hypothesis probability
was then selected as the global best fit.
From the spectral fits we computed
 the intrinsic (i.e.,~extinction corrected) X-ray fluxes  $F_{X,tc}$
for the total band ($0.5-8$~keV),
as derived from the spectral fit parameters,
and the corresponding X-ray luminosities $L_{X,tc}$, assuming a
source distance of 2.3~kpc.

The derived  plasma temperatures range from $\approx 3$~MK
up to $\approx 85$~MK (for a star that exhibited a strong X-ray 
flare during our observation).
The X-ray luminosities derived for the X-ray sources with 
infrared counterparts are
in the typical ranges found for young stars in other star forming regions
\citep[see, e.g.,][]{Preibisch_coup_orig}.

Owing to the moderate number of counts in most spectra,
the uncertainties of the model parameters are often rather high.
For many objects, the fitting analysis 
 did not yield a clear unique best-fit model;
two (or more) regions in the $N_{\rm H} - kT$ parameter plane
yielded very similar values for the best $\chi^2$.
This ambiguity in the best-fit parameters can lead to 
substantial uncertainties in the derived X-ray luminosities.
Furthermore, several
fits hit the lower $N_{\rm H}$ limit, suggesting
that the extinction is small, but leaving the value of $N_{\rm H}$ 
(and thus the un-absorbed X-ray luminosity) poorly constrained.
We therefore decided to use the spectral fitting
results only for those few sources for which reasonably well-defined best-fit
parameters could be determined. These cases are described individually
in the text below.

We note that
the brightest ACIS-I X-ray source (number 943; 190 net counts)
is a special case: it has no optical counterpart, only an extremely
faint infrared counterpart, and a very hard X-ray spectrum that can be
well fit by a power-law model.
As described in more detail in Appendix \ref{agn.sec}, 
these properties suggest it to be an extragalactic source, most likely an obscured quasar.

\subsection{X-ray fluxes for fainter sources}

Since the large majority of the detected X-ray sources is 
too faint for meaningful spectral fitting analysis,
we need another way to estimate the intrinsic X-ray luminosities.
In the course of the AE analysis, 
an estimate of the incident energy flux onto the telescope is calculated
for each source.
However, these flux values are known to be biased because their computation
is based on the (non-physical) assumption of a flat incident spectrum,
which is not correct for most X-ray sources.

For the X-ray sources that are identified with cluster stars,
we can improve the accuracy of the flux estimate by using
 the available a priori information on the typical
shape of X-ray spectra of young stars.
We can also use the
available constraints on the extinction (individual extinction estimates
available for the B-type stars, or an empirical mean value for the
extinction) and estimate not only the apparent, but also 
the intrinsic (i.e.,~absorption corrected)
X-ray fluxes and thus the X-ray luminosities of these stars.
For this, we employed the
\texttt{srcflux} tool\footnote{see 
\texttt{http://cxc.harvard.edu/ciao/ahelp/srcflux.html}}
in the \textit{Chandra Interactive Analysis of Observations}
(CIAO) software package.
This tool determines the conversion factor from count rate to 
apparent as well as unabsorbed energy flux
for a given spectral model, creating ARFs and RMFs for each source.
It is necessary to specify a model for the X-ray spectrum
and for the X-ray absorption caused by the ISM along the line of sight.
We used a thermal plasma spectrum 
with a plasma temperature of $kT = 1$~keV, which
is a typical value for young coronally active stars
\citep[see][]{Preibisch_coup_orig}.
For the absorption we used the model xsphabs and set the column
density to $N_{\mathrm{H}} = 0.15 \times 10^{22}\,\mathrm{cm}^{-2}$;
this corresponds to a visual extinction of 
$A_V \approx 0.75\,\mathrm{mag}$, which is the mean value of the
individual extinctions determined for the bright cluster stars.

%%%%%%%%%%%%%%%%%%%%%%%%%%%%%%%%%%%%%%%%%%%%%%%%%%%%%%%%%%%%%%%%%%%%%%%%%%

\section{Infrared and optical counterparts of the X-ray sources}

The identification of optical/infrared (OIR) counterparts of the 
X-ray sources is  required in order to obtain
essential information about the nature and
properties of the X-ray emitting objects, in particular to discern between
stellar X-ray sources  and (mostly extragalactic) contaminators.

The number of expected extragalactic contaminators can be estimated
by comparing the point-source sensitivity of our ACIS-I image
to the cumulative number counts of AGN found in the \textit{Chandra} Deep
Field South project \citep[see][]{Lehmer12}. 
This suggests 
$\approx 120 - 140$ AGN among the detected X-ray sources.
X-ray source 943, described
in Appendix \ref{agn.sec}, seems to be the X-ray brightest of these extragalactic 
X-ray sources.
A large fraction of the OIR counterparts of these 
extragalactic X-ray sources are expected to be too faint to be
detected in our OIR images.

Stellar X-ray sources, on the other hand, should have relatively bright
OIR counterparts.
 The expected optical and infrared 
magnitudes of young low-mass stars of NGC~3293  can be easily estimated.
Assuming  an age of about 10~Myr and a typical extinction of
$A_V \approx 1$~mag,
stars with masses of $[\,1.0\,, \,0.5\,, \,0.1\,]\,M_\odot$ are predicted to
have visual magnitudes of $V \approx [\,19.0\,, \,20.8\,, \,25.9\,]$
and NIR magnitudes of $H \approx [\,16.0\,, \,  16.4\,, \,  18.3\,]$
according to the pre-main-sequence stellar models of
\citet{Siess00}.
These numbers show that typical optical images are not sensitive
enough to detect the full low-mass population in NGC~3293.
In the NIR regime, these low-mass stars are much easier to detect, but
nevertheless too faint to be contained in all-sky catalogs such as the 2MASS data.

\subsection{VISTA near-infrared images and point source catalog}

As the main source for the identification and infrared characterization of the
X-ray sources we used the images and the point source photometric catalog from our
VISTA Carina Nebula Survey (VCNS), which is  described in detail in \citet{VISTA1}.
These VISTA data provide subarcsecond resolution and  $5 \sigma$  point-source sensitivities to $[J, H, K_s] \approx [20, 19.5, 18.5]$~mag.
Comparing these magnitude limits to the above mentioned expected magnitudes
of low-mass stars in NGC~3293 shows that the VISTA data
are clearly deep enough to detect {all low-mass members of the cluster},
i.e.,~down to $0.1\,M_\odot$, and even through several magnitudes of
visual extinction.
There are 35\,143 VISTA catalog sources with valid photometry in all three  NIR 
bands in the ACIS-I field of view. 
Five extremely bright stars in the cluster are so heavily saturated in the
VISTA images that they were not recognized as point sources in the VISTA
data processing; for these five stars (with 
magnitudes between $J = 3.5$  and $J = 8.5$),
positions and magnitudes from the 2MASS point source catalog were used.

Like any large-scale catalog, the VISTA catalog is not 100\% perfect and
complete. 
As described in Sect 3.2 of \citet{VISTA1}, 
some point-like objects, which are clearly visible in the VISTA images,
are missing from the catalog. This problem is mostly restricted to areas 
close to very bright stars where the wings of the point spread function
and numerous diffraction spikes produce a bright and spatially variable 
background pattern\footnote{The $\sim 1\arcmin$ area around the star V361 Car      
in the VISTA image shown in Fig.~\ref{H-cx.fig} provides a
good example of this effect.}, and also concerns close companions to some stars.
This sample of additional infrared sources is considered separately after the
matching with the VISTA catalog as described below 
in Sect. \ref{matching.sec}.

\subsection{Optical images and catalog from VST}

We used optical images of NGC~3293 that were obtained with Omegacam at the
ESO 2.6~m VLT Survey Telescope \citep[VST;][]{2011Msngr.146....8K}
in the $u$-, $g$-, $r$-, $i$-SDSS filters and an H$\alpha$ filter
as part of the VPHAS+ survey \citep{Drew14}.
The $5\sigma$ detection limits in these images are approximately
$u=20.5$,  $g=22.5$,  $r=21.4$, and  $i=20.5$ (all magnitudes in the
Vega system).

The ESO archive contains the fully reduced, calibrated,
 and verified Phase 3 data products
from this survey.
We downloaded all reduced images covering NGC~3293, and
also retrieved the VPHAS-DR2 Point Source Catalogue data for this
area, which provide band-merged PSF and aperture photometry 
for the point-like objects (``primary sources'' 
in the VPHAS+ terminology).
8581 of these VST primary sources are in the ACIS-I field of view.
The $g$-band magnitudes of these stars range from $g \approx 12.6$ (close to the
saturation limit) down to $g \sim 23$.
Comparing this faint magnitude limit to the above mentioned expected magnitudes
of low-mass stars in NGC~3293 shows that the VST data
should detect most lightly absorbed ($A_V \la 1$~mag) low-mass members 
of the cluster
down to $\ga 0.5\,M_\odot$, but will miss many of the lower-mass
members and also stars with visual extinction of more than about one visual
magnitude. The B-type and most A-type 
stars in NGC~3293 are also missing from the catalog, due to the saturation limit.
For these bright stars, optical photometry is available from \citet{Baume03} 
and \citet{Dufton06}.

We note that the VST catalog suffers from a similar   incompleteness 
to that of  our VISTA catalog, in the sense that 
some point-like objects that are 
clearly visible in the images are missing from the catalog.
Another limitation of the VST images is that many of the very bright stars
in the cluster center are strongly saturated in the CCD images.
The bleeding streaks caused by these saturated stars extend up to 1.5 arcmin 
in two directions from each
saturated source, and have widths up to about 20 pixels.
In the central $\sim 10$ square arcmin region, 
these bleeding streaks cover a significant fraction of the image area.
Faint sources may be completely hidden behind these bleeding streaks.
A visual inspection of the VST $g$-band image showed that at least 12
X-ray sources are located at positions within or just at the edge of such a 
bleeding streak or other image artifacts, which prevents the identification
of optical counterparts.
No attempt was made to construct a list of point sources missing from the
VST catalog; the matching analysis was restricted to the 
VPHAS catalog sources only.

\subsection{Further optical photometry and spectral data}

Our master catalog (based on the VISTA point source catalog)
was complemented by the 
optical (UBVRI H$\alpha$) photometry catalog for NGC~3293
from \citet{Baume03}, which provides
$V$-band magnitudes (down to $V \sim 20$)
for 1690 stars in the cluster and optical colors for a subset of them.

We also added information available in the literature about the spectral 
and other stellar parameters of stars in NGC~3293 into our
catalog.
Most of these data originate from the 
VLT-FLAMES multi-object spectroscopic survey of massive stars in NGC~3293
\citep{Evans05,Dufton06}, 
which provided spectral types for 131 stars and luminosities for 92 of these.
In the ACIS-I field of view,  our final catalog lists
97 B-type stars, 17 A-type stars, 2 F-type stars,
and 1 M-type star (the M1.5~Iab supergiant V361~Car).

\begin{figure*}
\centering
\includegraphics[width=18cm]{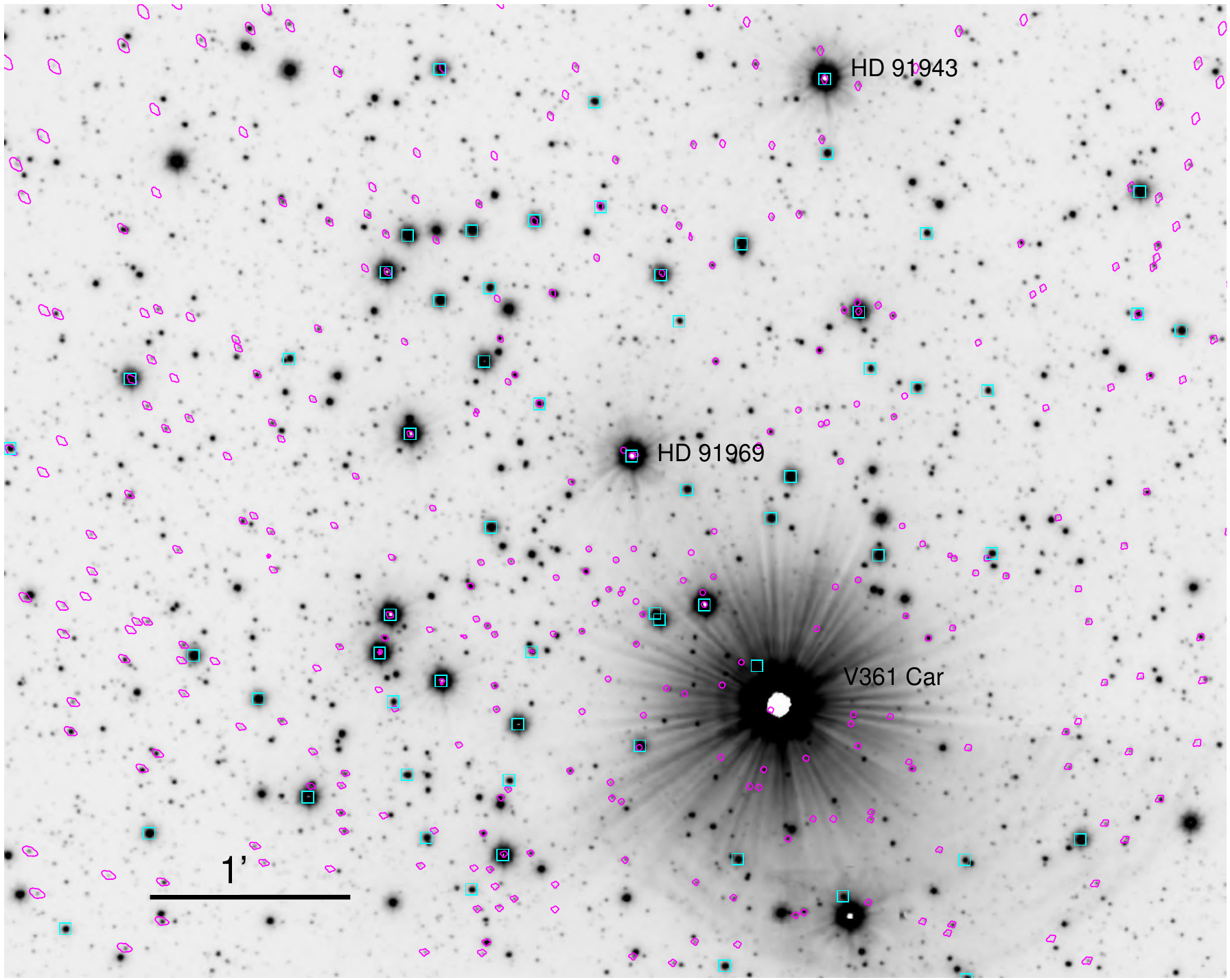}
\caption{VISTA $H$-band image of the central part of NGC~3293.
North is up and east to the left.
The known B-type stars are marked by the cyan squares.
The \textit{Chandra} X-ray sources are marked by their individual
source extraction polygons drawn in magenta.
The M1.5Iab supergiant V361~Car, the 
B0~Iab star HD~91969, and the B0.7~Ib star HD~91943 are labeled.
\label{H-cx.fig}}
\end{figure*}

%%%%%%%%%%%%%%%%%%%%%%%%%%%%%%%%%%%%%%%%%%%%%%%%%%%%%%%%%%%%%%%%%%%%%%%%%%%%%%%%%%

\subsection{Matching of the X-ray sources with the optical/infrared catalogs \label{matching.sec}}

Figure~\ref{H-cx.fig} shows a NIR image of the central part of 
the cluster with the polygons of the X-ray source extraction regions
overlaid. The high density of point-like sources in the NIR image
(which is largely due to the strong galactic background
at the cluster's position very close to the galactic plane)
shows that a proper identification of the counterparts to our
X-ray sources requires a careful and conservative matching procedure
in order to avoid false matches with unrelated background objects.

The identification of infrared and optical counterparts to 
the X-ray sources was
performed with a two-step procedure: an automatic source matching
based on catalog coordinates, followed by a
detailed individual inspection
in order to resolve problematic cases such as
multiple possible matches or objects missing from the optical/NIR catalogs.

For the first step, we employed the IDL tool\footnote{see\, 
{\tt www2.astro.psu.edu/xray/docs/TARA/TARA\_users\_guide/node11.html}}
{\tt match\_xy}, as described in \citet{CCCP-catalog}.
The maximum acceptable matching separation between an X-ray source and 
a counterpart 
was based on the individual source position errors 
assuming Gaussian 
distributions, scaled so that $\sim 99\%$ of the true associations 
should be identified as matches.
For the X-ray sources, the positional uncertainties were determined by
AE, and for the
VISTA and VST sources we assumed position uncertainties of
$0.1''$.
The algorithm in {\tt match\_xy} lists the 
 most significant match of each X-ray source as its 
``Primary Match'', while  any other possible
significant matches, if present, are considered  ``Secondary Matches''.
In a second stage, the algorithm resolves possible 
many-to-one and one-to-many relationships between the X-ray 
and infrared/optical catalogs.
Clear one-to-one relationships are classified as ``successful primary matches'',
while in cases where, e.g., two X-ray sources are significantly close to
a single infrared source, the less significant primary match
 is labeled as ``failed''.
This finally provided a reasonable one-to-one set of matches, consisting
of 695 successful primary matches to
VISTA catalog sources in the ACIS-I field.

For 42 X-ray sources, one or more successful secondary matches
were identified by {\tt match\_xy} in the VISTA catalog. 
These cases require further inspection since 
 it is not guaranteed
that the closest match is always the correct physical counterpart, due to the high surface density of infrared sources in our deep
VISTA images.
Instead, a physically unrelated (e.g., background) infrared source
may appear in the matching region just by chance
and produce a ``false match'', and might even degrade the 
true infrared counterpart to a secondary match.
A priori information about the typical infrared properties
of X-ray detected young stars can be used to identify 
possible cases where this problem occurs.
Since we observe a young stellar cluster and 
know from the  X-ray detection limit that most of the
X-ray detected objects should be young stars with
 masses of $\ga 0.5\,M_\odot$,
these stars should typically be relatively bright NIR sources.
Therefore, all cases where an X-ray source has an unexpectedly faint 
primary match
and a considerably brighter secondary match (i.e.,~with the magnitudes
 expected for a young star) deserve special attention.
We found eight such cases, and 
replaced the original faint primary match by the considerably 
brighter secondary match.

As the last step in the source matching procedure,
we finally considered the above mentioned list of point-like sources 
identified in our visual inspection of the VISTA images but missing from the  VISTA source catalog.
The locations of these additional VISTA sources were compared to the 
X-ray source locations for which
{\tt match\_xy} did not report a match with a VISTA catalog star.
In this way, 61 additional infrared point source matches to
\textit{Chandra} sources were identified.
We note that most of these 61 additional infrared sources have
discernible counterparts in the optical VST images, and 20 of them
are even listed as point sources in the VST catalog, demonstrating
the reality of these infrared sources.

Since most of these objects
are affected by a spatially variable background or by partial blending with
other nearby sources in the VISTA images, measurements of their photometry 
can suffer from relatively large
uncertainties. In order to quantify these uncertainties,
we performed aperture photometric measurements in the VISTA images, employing 
different values for the sizes of the background regions, and determined the
scatter of the resulting aperture flux measurements for each source.
For 21 of these sources, the relative scatter of these flux measurements
exceeded 20\%; these sources were completely excluded from all
further analysis steps.
For the 40 sources for which the relative scatter of the flux measurements
was less than 20\%, we determined their magnitudes, but
these sources were {not} included
in the sample of VISTA catalog sources; they are considered 
in some of the analysis steps below, but are always
clearly kept separate from the VISTA catalog sources.
There is thus no danger that these additional sources will contaminate our
results. 
\bigskip

In the matching with the VST optical catalog, we considered
all sources that are listed as primary source in the
VPHAS-DR2 catalog.
The matching with the VST catalog resulted
in 496 successful primary matches in the ACIS-I field. 
In the course of our visual inspection of the X-ray source positions
in the VST images, we found one case where an image artifact that was
erroneously listed as point source in the VPHAS-DR2 catalog
coincided with an X-ray source. More details about this artifact
are given in Appendix \ref{false-match.sec}.

The final results of the matching procedure can be summarized as follows:
756,  i.e.,~74.0 \% of all X-ray sources in the ACIS-I field,
 have a match with a VISTA source (695 of them, i.e.,~68.0\%, with a 
VISTA catalog source). 
For the VST catalog, we have 491 matches of VST catalog primary source 
to ACIS-I X-ray sources after removing the above mentioned artifact,
and four cases where
the VST match was not identical to the VISTA match; this yields a 
match fraction of 48.0\%.
The lower rate of counterparts in the VST catalog is easily
understood as a consequence of the different sensitivity limits of 
the two catalogs.
While the VISTA catalog is deep enough to detect $0.1\,M_\odot$ stars
in NGC~3293
with extinctions up to $A_V \simeq 5$~mag,
the sensitivity limit of the VST catalog  corresponds to
$1\,M_\odot$ stars with extinctions up to $A_V \simeq 3.5$~mag,
or $0.5\,M_\odot$ stars with extinctions up to $A_V \simeq 2$~mag.
Therefore, most of the $<0.5\,M_\odot$ stars and a substantial fraction
of the $[0.5-1]\,M_\odot$ stars remain undetected
and will be missing from the VST catalog.

Information about the VISTA and VST matches to the 
individual X-ray sources 
is contained in the electronic Table 2, available at the CDS.

Finally, we can compare the fraction of IR matches for the two above-mentioned subsamples of our X-ray catalog.
In the sample of 849 highly reliable ACIS-I sources, we find
593 objects (i.e.,~$69.8 \pm 0.2\%$) that have a match with a VISTA catalog source.
In the less reliable sample, 102 of the 173 ACIS-I sources (i.e.,~$59.0 \pm 3.7\%$)
 have a VISTA catalog match. 
The lower match fraction for the less reliable subsample 
is partly due
to a higher fraction of spurious sources among the
less reliable sources.
It should be noted, however,  that
X-ray sources  detected with lower significance  generally have weaker X-ray fluxes.
Since  there is some
correlation between X-ray flux and infrared brightness for most classes of X-ray emitting objects, 
the less significant true X-ray sources are also less likely to have an infrared
counterpart that is detected in the available infrared images.

\subsection{Reliability of X-ray sources and their infrared matches \label{reliability.sec}}

In order to determine the reliability of the following analysis,
we briefly discuss a few fundamental aspects of the sample of
X-ray sources and their OIR matches.
In general, any X-ray source list is composed of {real sources} and
{spurious sources}.
Concerning the identification of OIR matches to the X-ray sources,
it is necessary to take into account that
brighter sources are usually detected with higher
significance, i.e.,~there is a correlation between X-ray source brightness and significance.
There is also a correlation between X-ray flux and OIR brightness
for most classes of X-ray emitting objects; therefore,
the less significant true X-ray sources are, on average, fainter in the OIR range and
therefore also less likely to have a
counterpart that is detected in the available OIR images.

For the OIR counterparts of the real X-ray sources
there are two possibilities:
some of the X-ray sources can be expected to have a counterpart
bright enough to be detected in the available OIR images
(in the present study, this concerns all young stars in NGC~3293 in 
the deep VISTA images, and the $\ga 1\,M_\odot$ stars in the VST images).
Another fraction of the X-ray sources will be extragalactic objects 
(mainly AGN), most of which should be very faint at OIR
wavelengths and thus remain undetected in the available OIR images.

If we now consider the possible results of the X-ray to OIR
source matching, we have to distinguish four possibilities for
the real X-ray sources:
(a) \emph{correct positive matches}, when the physical association between an
X-ray source and the corresponding OIR source is correctly identified;
(b) \emph{correct negative matches}, when an X-ray source is not matched
to any OIR catalog source because  the OIR counterpart is too faint to 
be detected in the OIR images;
(c) \emph{false negative matches}, when the true OIR counterpart is not 
considered  a match, e.g.,~because the angular distance of the catalog
entry is wider than the matching limit; and 
(d) \emph{false positive matches}, when unassociated objects are incorrectly considered
 a match, e.g.,~because the catalog coordinates of some unrelated
 OIR source are by chance closer to the X-ray source position than 
the coordinates of the true counterpart, or because the true counterpart is
undetected in the available OIR images and an unrelated OIR source
is close enough to the X-ray source position.

For the spurious X-ray sources, there are two possibilities:
(e) \emph{correct negative matches}, when no OIR counterpart is
associated with the non-existing X-ray source, and
(f) \emph{false positive matches}, when an unrelated OIR source
is wrongly considered to be  a match to the non-existing X-ray source.

In any matching procedure, some level of spurious matching will inevitably occur.
These issues were discussed in detail in \citet{CCCP-catalog},
where quantitative estimates for the expected numbers of 
false positive matches with random unrelated infrared sources
were derived for the case of the CCCP data set.
However, the numbers for the expected fractions of false positive matches 
derived by \citet{CCCP-catalog} cannot be directly applied to the present study because  the
underlying infrared catalogs are quite different.
The highest estimated false match fraction derived in \citet{CCCP-catalog} 
was based on the
matching of the CCCP X-ray source list with the very deep HAWK-I infrared
catalog \citep[see][]{HAWKI-survey}. Since the HAWK-I catalog 
is about 2 magnitudes deeper 
(i.e.,~a factor of $\ga 6$)  than the VISTA catalog
we use here,  it 
contains a large number of very faint (mostly background)
objects that provide numerous possibilities for random false positive matches;
however, these very faint objects remain undetected in the VISTA catalog
and thus cannot produce random false positive matches in our case.
Since the HAWK-I catalog lists on average $\approx 470$ IR sources per 
square-arcminute in the CCCP field,
while the VISTA catalog contains  only 140 sources per square arcminute in the
NGC~3293 region, the probability of obtaining a random 
false positive match with the VISTA catalog is at least about 3 times lower
than reported for the HAWK-I catalog in the CCCP field.

In order to derive a quantitative estimate for the occurrence rate of 
random false positive matches for X-ray sources in our NGC~3293 data set,
we performed a set of random matching simulations that are described in 
Appendix \ref{random-matches.sec}.
These simulations show that 
the average probability that a spurious X-ray source will get a false positive
VISTA catalog match is about 23\%.
It is very important to note that this is {not} equal to the
expected fraction of false matches in our sample since in reality
a large fraction of our X-ray sources are true sources that have a
counterpart in one of the young stars in NGC~3293;
the 23\% probability for false positive matches applies only to the subset of
X-ray sources that have no true physical counterparts in our VISTA images, 
which is considerably smaller than the full sample.

In order to estimate the possible consequences of such false positive matches 
on the results of our study of the cluster properties described below,
it is important to note that  
the large majority of all random false positive matches 
are with very faint IR sources. This is just a consequence of the fact that the number of
IR sources increases strongly when going towards fainter magnitudes.
The majority of the VISTA catalog sources in the NGC~3293 field have magnitudes $J > 18.5$.
False matches with such very faint IR objects will only have  very 
minor effects on our analysis of the
color magnitude diagram presented below; 
e.g., such very faint IR objects
 cannot be confused with $M \ge 1\,M_\odot$ cluster stars,
which are considerably brighter.
Most of our X-ray sources with VISTA catalog matches are much brighter
and have magnitudes $J < 16.5$.
As shown in Appendix \ref{random-matches.sec}, the probability of a random
X-ray source obtaining a false positive match with a VISTA catalog source
brighter than $J = 17$ [$J = 16$] is just 4.6\% [2.5\%].
\medskip

If we assume that $\approx 20\%$, i.e.,~$\approx 200$
 of the sources, in our X-ray catalog  have no physical counterparts in the VISTA images
(i.e., they might be spurious sources),
we  
expect there will be $\approx 46$ false positive matches with VISTA catalog sources.
However, only $\approx 5- 9$ of these would be false positive matches to stars
that are bright enough ($J \leq 16 - 17$) to possibly affect
our star counts for the investigation of the cluster IMF
presented in Sect.~\ref{IMF_counts.sec}.

%%%%%%%%%%%%%%%%%%%%%%%%%%%%%%%%%%%%%%%%%%%%%%%%%%%%%%%%%%%%%%%%%%%%%%

\section{Exploring the cluster properties}

Our X-ray selected sample provides the first opportunity to study 
individually identified low-mass stars in NGC~3293.
We first consider the X-ray-to-infrared flux ratios of the objects,
and then
use color magnitude diagrams (CMDs) 
to derive information about the ages and masses of the
X-ray detected stars and information about the size of the
low-mass star population in the cluster.

%%%%%%%%%%%%%%%%%%%%%%%%%%%%%%%%%%%%%%%%%%%%%%%%%%%%%%%%%%%%%%%%%

\subsection{X-ray and infrared fluxes}

\begin{figure} \centering
\includegraphics[width=8.5cm]{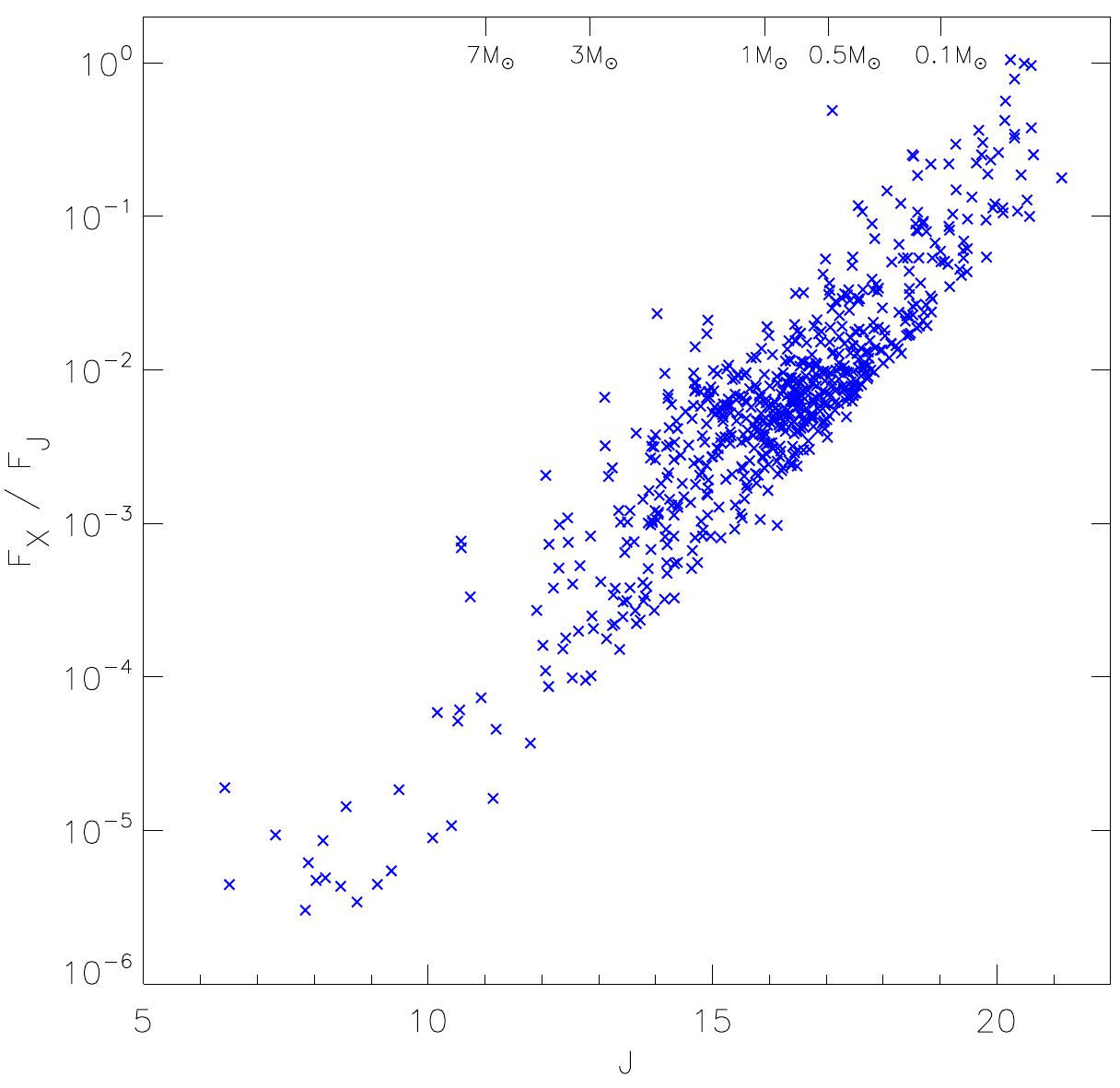} % }
\caption{Ratio of observed X-ray fluxes to observed $J$-band fluxes
 plotted against the $J$-band magnitudes of the X-ray sources
with infrared counterparts.
On the upper x-axis, the expected $J$-band magnitudes for 
10~Myr old stars with different masses are indicated according to the 
models of \citet{Siess00}.
\label{fxfj.fig}}
\end{figure}

Reliable bolometric luminosities (i.e.,~determined from spectroscopic information)
are only available for a relatively small number of stars in NGC~3293
(essentially the B-type stars, which we  discuss in Sect.~5).
For the bulk of the stellar population of NGC~3293, we thus cannot directly determine 
the ratio of X-ray to bolometric luminosity,
which would be a good diagnostic of the nature of X-ray emission.
Therefore, we employ here 
the observed $J$-band flux as a proxy for the stellar luminosity.
For late-type stars, the $J$-band flux is roughly proportional
to the stellar bolometric luminosity; furthermore, the influence of extinction is rather
small in the $J$-band ($A_J \approx 0.28 \times A_V$).

In Fig.~\ref{fxfj.fig} we show the ratio of observed X-ray and infrared fluxes
for the sources in NGC~3293, which shows that the majority of X-ray detected sources are objects for which
their $J$-band magnitudes suggest stellar masses in the range $\sim [2 - 0.5]\,M_\odot$,
consistent with the X-ray sensitivity limit. Most of the stars 
in the range $J \approx [14-18]$ show ratios around a typical value of
$\log \left( F_{\rm X}/F_J \right) \approx -2.2$.
Since for these low-mass stars, the bolometric flux is roughly a factor of $\sim 10$ higher 
than the $J$-band flux, this corresponds to fractional X-ray luminosities of
$\log \left( L_{\rm X}/L_{\rm bol} \right) \approx -3.2$, which is the typical
ratio for coronally active young low-mass stars.

The bright stars  in the range $J \approx [6-9]$ 
show typical ratios $\log \left( F_{\rm X}/F_J \right) \approx -5.2$.
Since for early B-type stars, the bolometric flux is typically a factor of $\sim 1000$ higher 
than the $J$-band flux, this translates roughly into
$\log \left( L_{\rm X}/L_{\rm bol} \right) \approx -8.2$. Again, this is very  
consistent with the results from other studies \citep[e.g.,][]{Stelzer05,Gagne11}.
A more accurate assessment of the X-ray properties of the B-type stars
in NGC~3293 is  given in Sect.~\ref{Bstar.sect}.

Some of the faintest infrared sources ($J \ga 19$) show
very high X-ray-to-infrared flux ratios; many of these objects
are probably extragalactic sources.

%%%%%%%%%%%%%%%%%%%%%%%%%%%%%%%%%%%%%%%%%%%%%%%%%%%%%%%%%%%%%%%%%

\subsection{Spatial distribution of the X-ray sources \label{spatial-distr.sec}}

\begin{figure} \centering
\includegraphics[width=8.5cm]{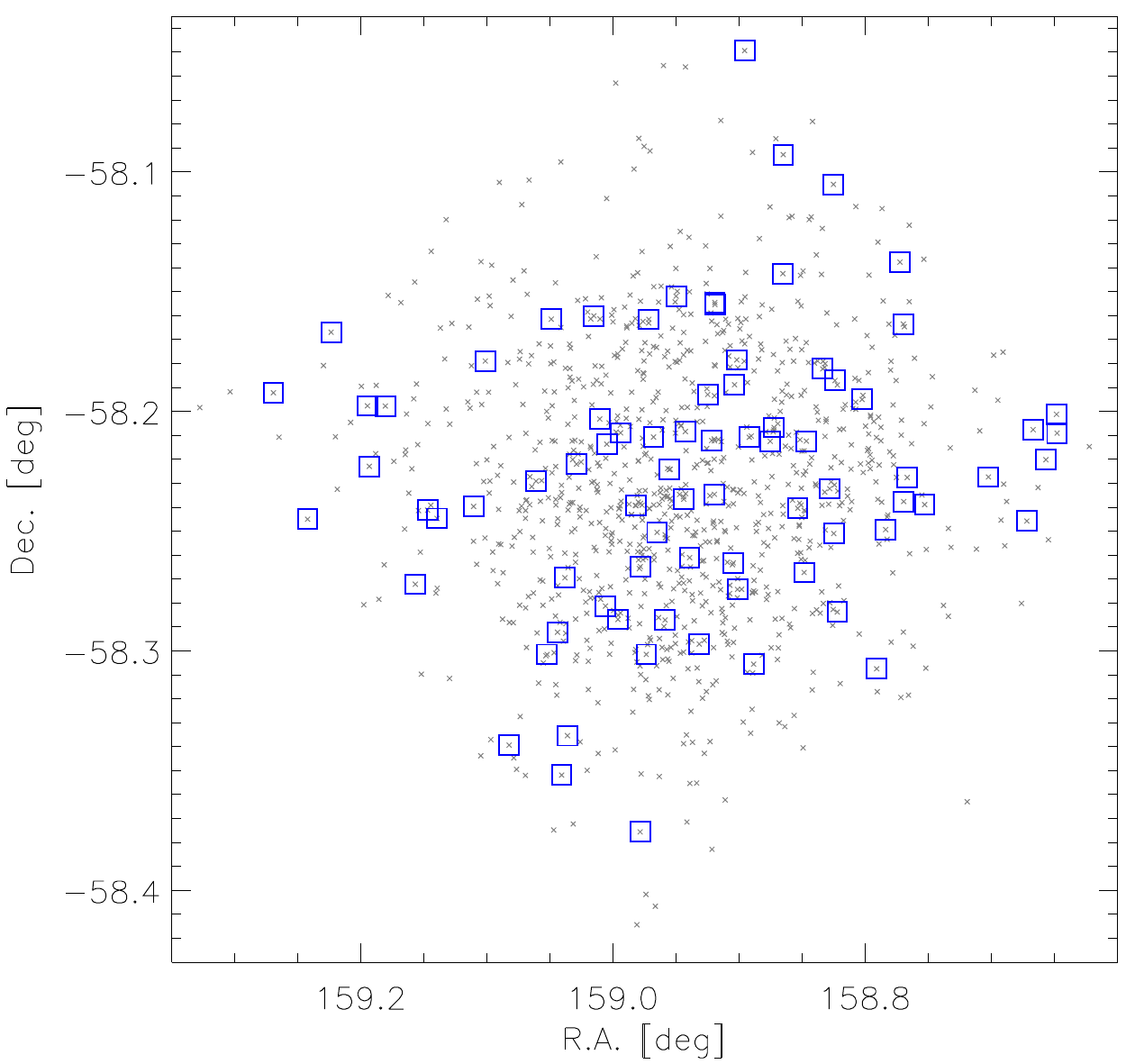} % }
\caption{Spatial distribution of the X-ray sources.
The small gray crosses show all X-ray sources. The blue squares
mark the spatially complete sample of  sources with an
observed X-ray photon flux
$\log F_{t,\mathrm{photon}} > -5.9\,\mathrm{photons}\,\mathrm{s}^{-1} \,\mathrm{cm}^{-2}$
that have VISTA counterparts with $J < 18$.
\label{map_complete.fig}}
\end{figure}

\begin{figure}\centering
\includegraphics[width=8.5cm]{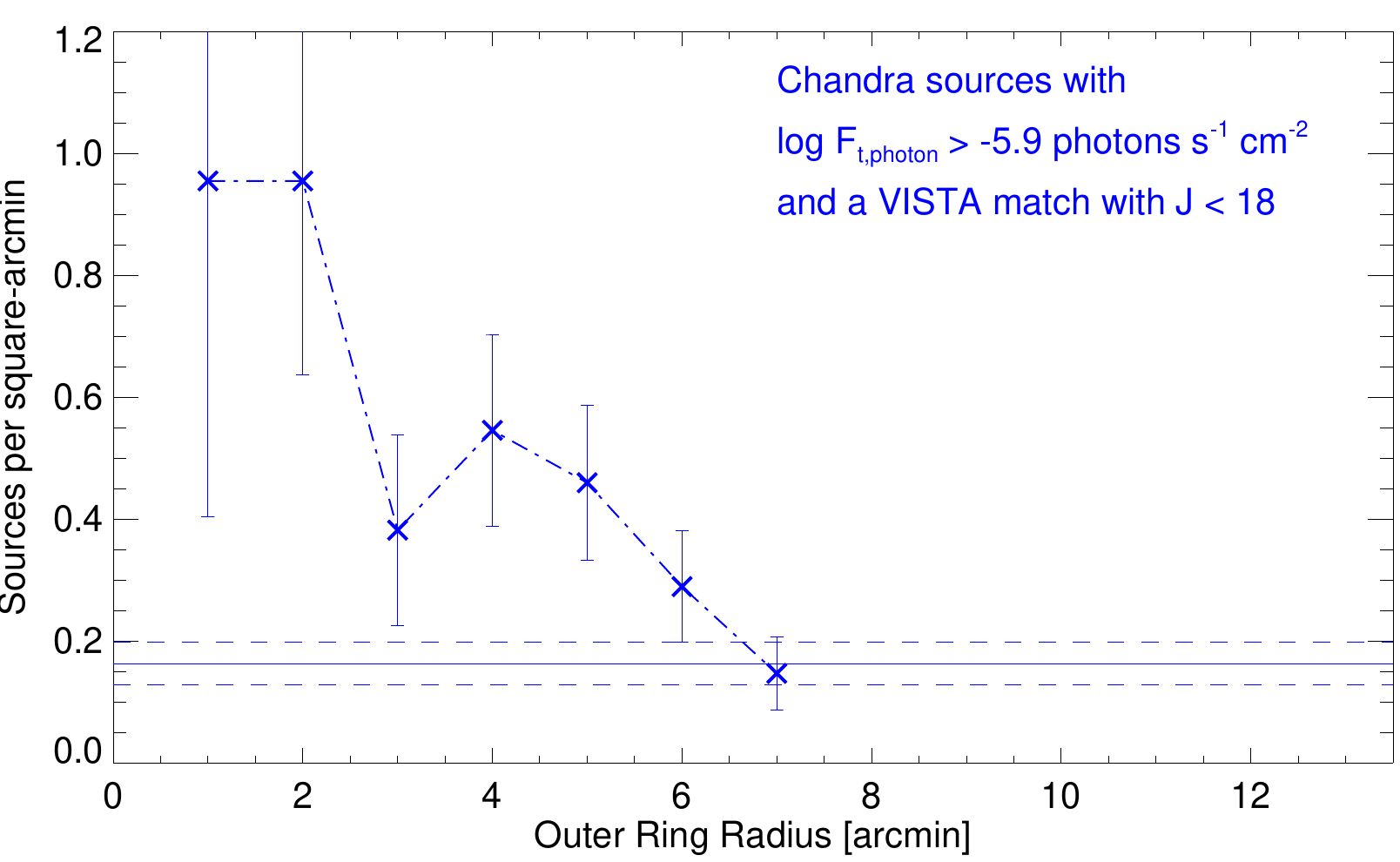}
\includegraphics[width=8.5cm]{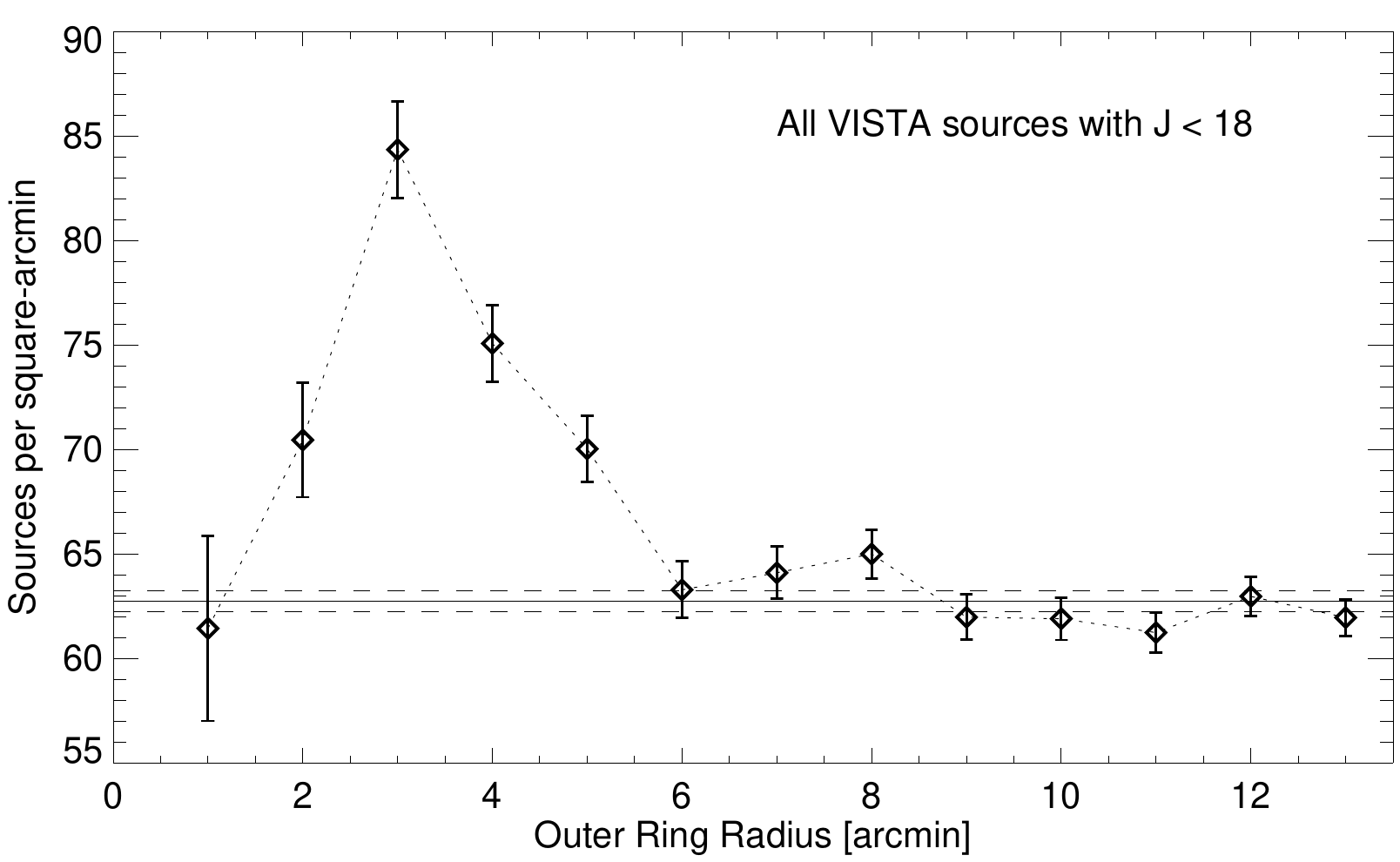}
\caption{Top: Radial profile of the surface density of
\textit{Chandra} sources in the
spatially complete sample that have a VISTA match with $J < 18$
in different rings around the optical cluster center.
The horizontal blue line and the parallel dashed lines show the
mean value and the uncertainty of the source density outside the 7~arcmin
region.
Bottom: Radial profile of the surface density of  all VISTA source with
$J < 18$ in different rings around the optical cluster center.
The horizontal line and the parallel dashed lines show the
mean value and the uncertainty of the source density in several comparison fields.
\label{radial-profiles.fig}}
\end{figure}

Next we consider the spatial  distribution of the X-ray sources with the aim
to derive information about the cluster size.
\citet{Slawson07} estimated the coronal radius of NGC~3293 to be
$\sim 5.5'$, but since this was based on optical data
that are not sensitive enough to detect the full cluster population
(especially in regions where the extinction is more than
$A_V \approx 1$~mag), an independent check is useful.
We can address the question regarding the cluster size with our
X-ray sample and with the deep VISTA images.

Considering the \textit{Chandra} data, it should be noted
that the sensitivity level of the observation is not uniform across the
observed field since the point-source sensitivity drops significantly 
with off-axis angle.
The most important factors that play a role here are
the mirror vignetting and the degradation of the point spread function,
both of which reduce the local sensitivity with increasing
off-axis angle. 
Following the strategy of \citet{CCCP-Clusters},
a ``spatially complete sample'' of X-ray sources can be
constructed by employing a limit to the observed photon flux of the
X-ray sources that is high enough to make sure that sources above this limit
can be detected in the full ACIS-I field of view, i.e.,~even at the edge of the
X-ray image. 
This threshold on the observed X-ray photon flux is
$\log F_{t,\mathrm{photon}} > -5.9\,\mathrm{photons}\,\mathrm{s}^{-1} \,\mathrm{cm}^{-2}$.

Since we intend to trace the spatial structure of the stellar cluster, we
use only those objects in this spatially complete X-ray sample that have a
match with a VISTA source with $J < 18$, as expected for the
large majority of X-ray detected young stars in NGC~3293.
This final sample contains 81 sources.
A map of the positions of these sources is shown in Fig.~\ref{map_complete.fig}.

We determined the surface-density of these X-ray sources as a function of
distance from the optical cluster center by counting the number of objects
in concentric annular regions with different radii. 
The resulting radial profile is shown in the upper part of
Fig.~\ref{radial-profiles.fig}. The source density in the center is more than
5 times higher than in the outer parts of the ACIS-I image, drops with
angular distance from the cluster center, and merges to the background value
at an angular distance of $\approx 6\!-\!7$~arcmin from the center.
This profile is  consistent with the above mentioned previous estimate of the
cluster radius and confirms that the
ACIS-I field of view covers the full area of the cluster.

The question  about the  spatial extent of the cluster can also be addressed with
the VISTA data. These data have several advantages.
First, they cover a considerably larger area around NGC~3293
(the angular distance from the cluster center to the nearest edge of the 
VISTA image is 12 arcmin). This makes it easy to define background comparison regions 
at angular distances of $\ga 10$ arcminutes for a reliable computation
of the background source density.
Second, the VISTA images do not suffer significantly from PSF degradation
and mirror vignetting, i.e.,~they provide a more homogeneous coverage of the
observed area.
Third, the VISTA data are sensitive  enough to
detect {all} cluster stars (down to $0.1\,M_\odot$, and even through substantial extinction
of $A_V \approx 5$~mag) and thus provide much larger source numbers and 
consequently better statistics in the source counts. 
However, there is also one disadvantage: several of the very bright stars
in the cluster are saturated, and their PSF-wings and diffraction spikes
create a high and complicated background halo around them, 
which severely restricts the detection of faint sources.

The radial profile of the surface density of all VISTA sources with $J < 18$
is shown in the lower part of Fig.~\ref{radial-profiles.fig}.
A clearly enhanced source density can be seen for radial distances between
$2$~arcmin and $6\!-\!8$~arcmin from the cluster center.
The low density in the center is caused by the strongly reduced detection efficiency
for faint sources near the very bright stars in the cluster center.
The VISTA density profile suggests a cluster radius in the range
$6\!-\!8$~arcmin.

Considering all these numbers together, a good estimate for the
angular size of the cluster is $\approx 7$~arcmin, which corresponds to
$\approx 4.7$~pc at 2.3~kpc distance.

%%%%%%%%%%%%%%%%%%%%%%%%%%%%%%%%%%%%%%%%%%%%%%%%%%%%%%%%%%%%%%%%%

\subsection{Analysis of the  color magnitude diagrams}

 In order to obtain information on the stellar properties
of the X-ray detected stars, we constructed color magnitude diagrams (CMDs)
and compared the location of the X-ray selected objects to the recent PARSEC
stellar evolution models described in \citet{PARSEC}.

\subsubsection{Near-infrared color magnitude diagram}

\begin{figure}
\includegraphics[width=9.0cm]{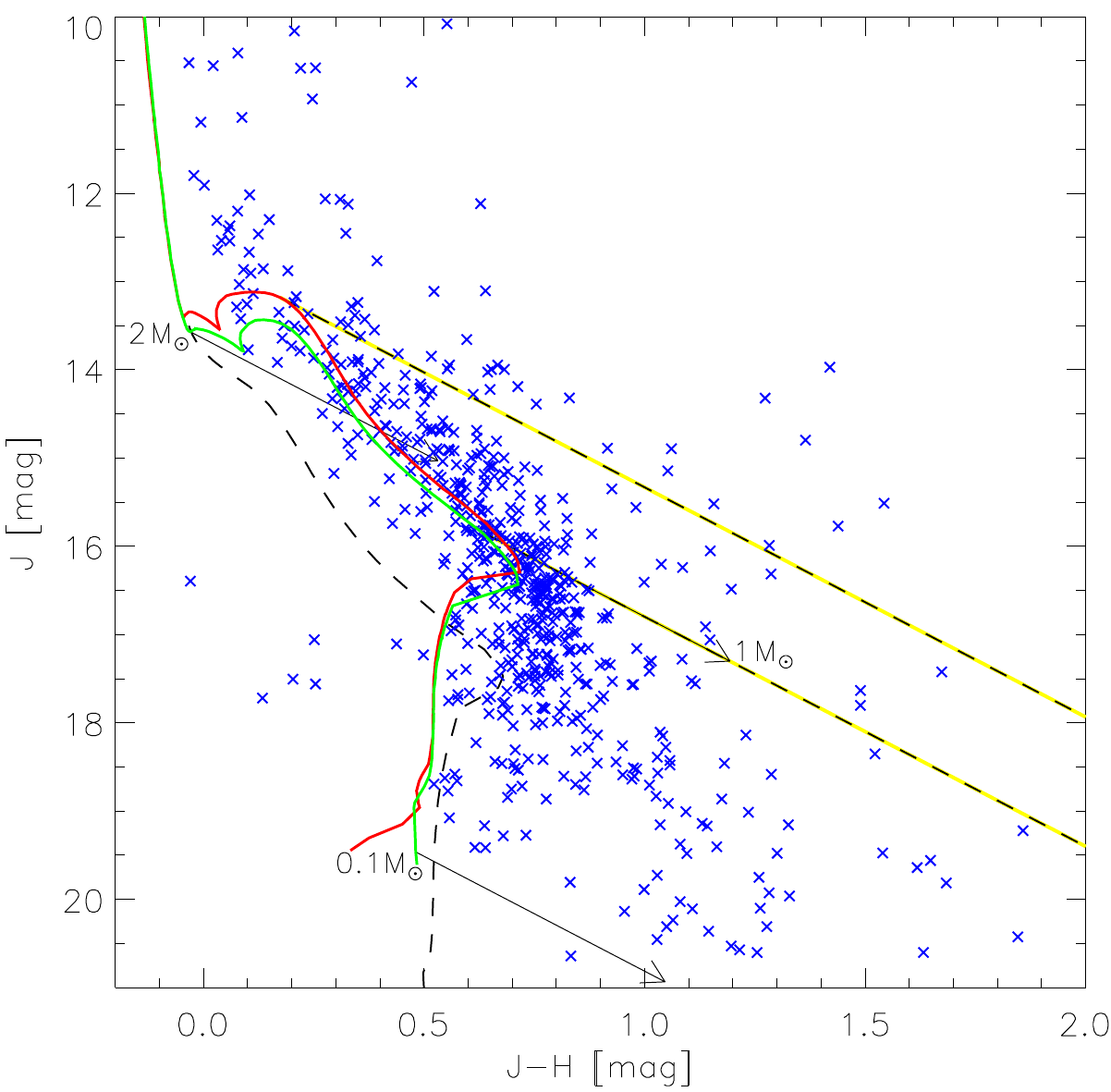} % }
\caption{
Near-infrared color magnitude diagram of the X-ray detected objects
in NGC~3293 (blue crosses).
The red and green solid lines show isochrones for  stellar ages of 8~Myr and
10~Myr, based on the PARSEC stellar models \citep{PARSEC}.
The black dashed line indicates the main sequence.
The arrows indicate reddening vectors for $A_V = 5$~mag starting at the
location of 10~Myr old stars with masses of $2\,M_\odot$, $1\,M_\odot$, and
$0.1\,M_\odot$. The yellow-and-black dashed lines show the upper and lower limits
for the ranges that were used for counting the number of stars in the $[1-2]\,M_\odot$
interval.
\label{cmd.fig}}
\end{figure}

\begin{figure}\centering
\includegraphics[width=9.0cm]{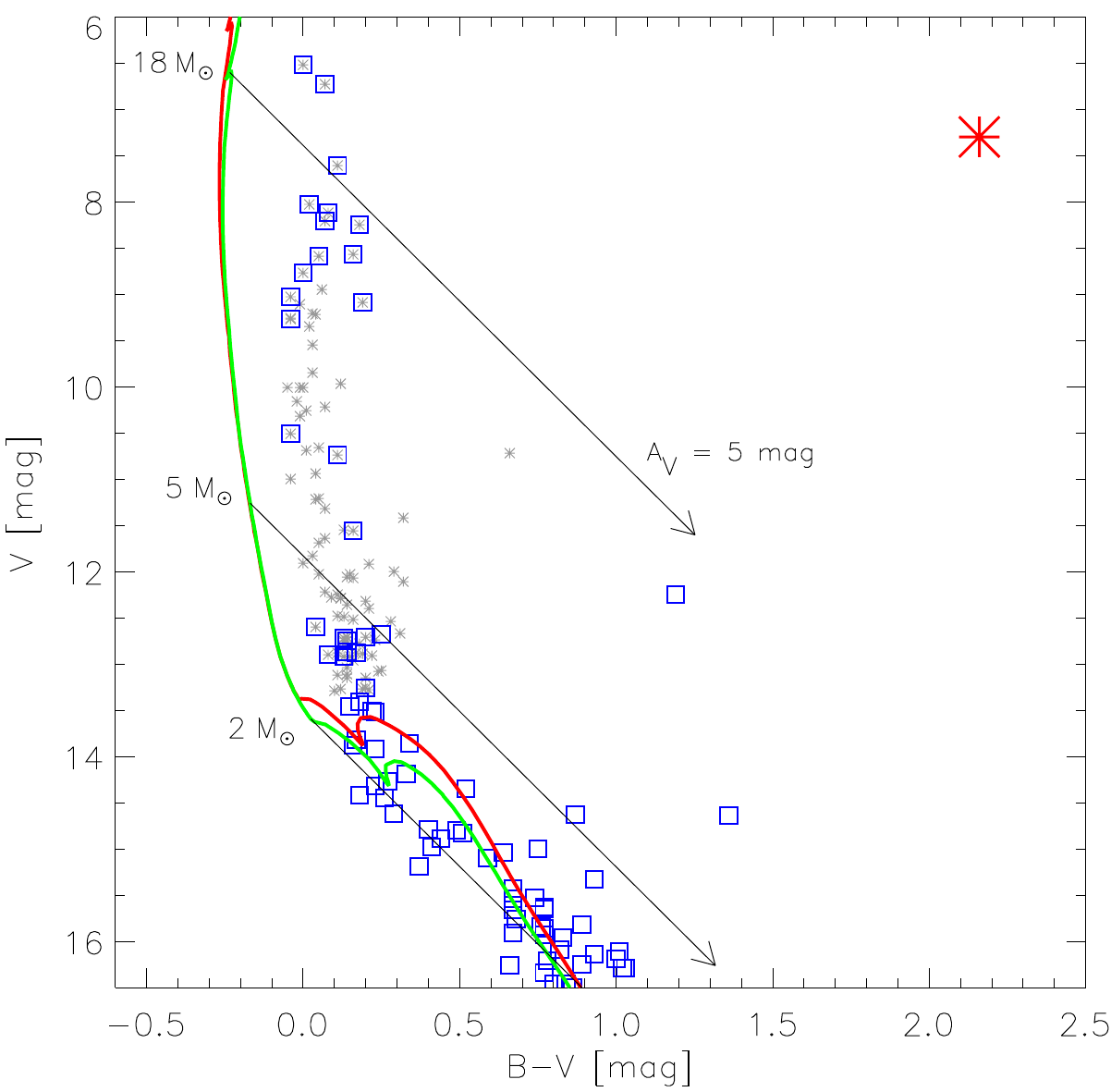}
\includegraphics[width=9.0cm]{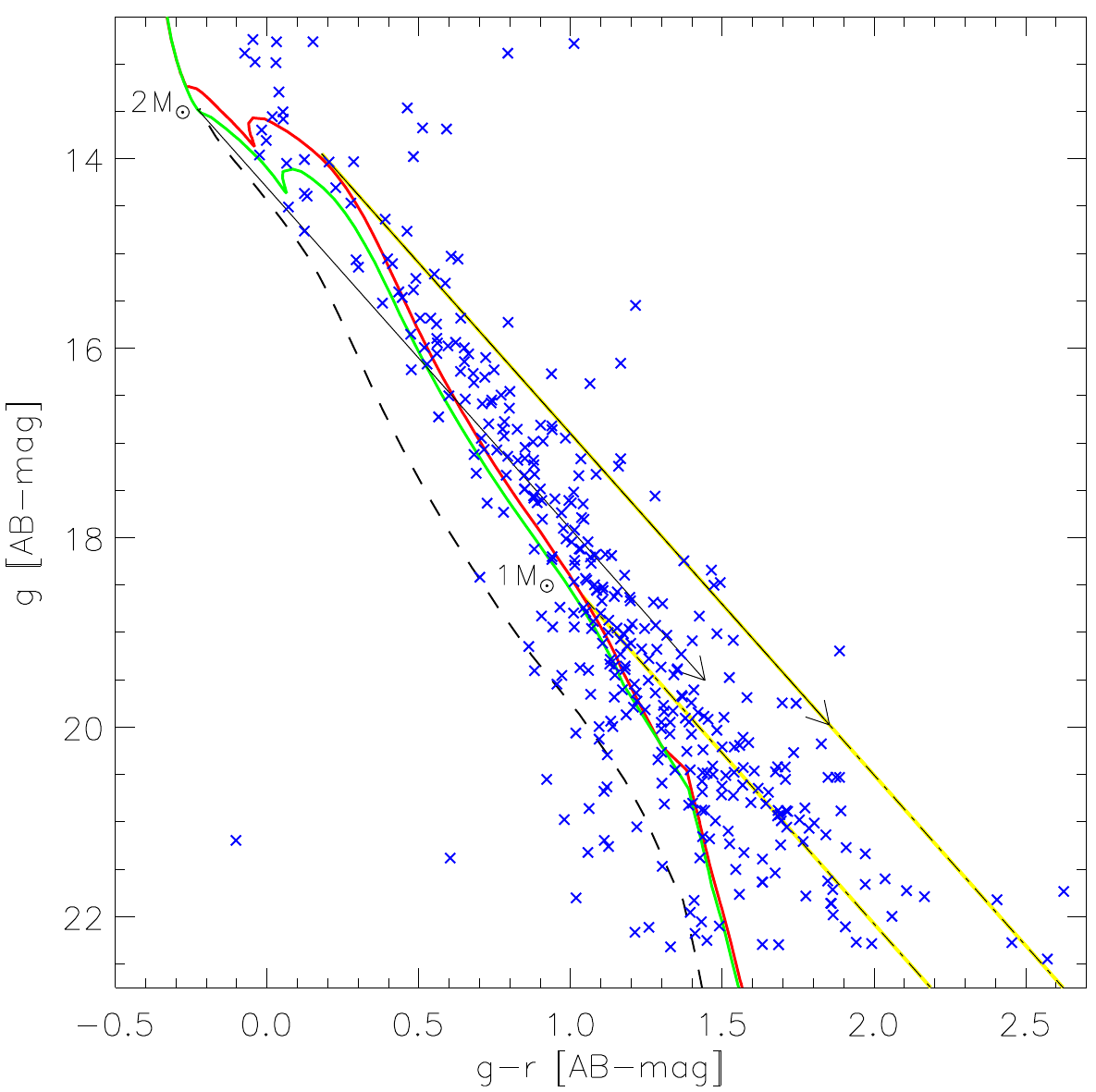}
\caption{Top:
Optical color magnitude diagram of the X-ray detected objects in NGC~3293 based on the optical
photometry from \citet{Baume03}, complemented by values from \citet{Dufton06} for
the B-type stars. The gray asterisks mark stars with known spectral type in the ACIS-I field.
The X-ray sources are marked by blue squares.
The  solid lines in red and green show isochrones for  stellar ages of 8~Myr and
10~Myr based on the PARSEC stellar models. 
The arrows indicate reddening vectors for $A_V = 5$~mag starting at the
location of 10~Myr old stars with masses of $18\,M_\odot$, $5\,M_\odot$, and
$2\,M_\odot$.
The large red star marks the  M1.5Iab supergiant V361~Car.\newline
Bottom: Optical color magnitude diagram of the X-ray detected stars in NGC~3293 based on photometry
from the VST VPHAS-DR2 survey data (blue crosses).
The  solid lines in red and green show isochrones for stellar ages of 8~Myr and
10~Myr based on the PARSEC stellar models,  and the black dashed line
indicates the main sequence.
The arrows show reddening vectors for $A_V = 5$~mag starting at the
location of 10~Myr old stars with masses of $2\,M_\odot$ and $1\,M_\odot$.
The yellow-and-black dot-dashed lines show the upper and lower limits
for the ranges that were used to count the number of stars in the $[1-2]\,M_\odot$
interval.
\label{cmd-opt.fig}}
\end{figure}

In Fig.~\ref{cmd.fig} we show the $J$ versus $J-H$  
color magnitude diagram of the X-ray detected objects in NGC~3293.
The range of $J$-magnitudes is restricted to the fainter values
in order to show the location of the low-mass stars more clearly;
the color magnitude locations of the brightest (i.e.,~the early-type) stars
can be seen in Fig.~\ref{cmd-opt.fig} and are discussed below.

The large majority of the X-ray selected objects 
are at CMD locations corresponding to stellar masses in the
$[0.5\!-\!2]\,M_\odot$ range,
as expected from the X-ray detection limits, and extinctions of a few visual magnitudes. 

At the bottom of the CMD ($J \ga 19$), the X-ray selected objects
show a wide range of colors consistent with background objects and
the expected locus of extragalactic sources.

\subsubsection{Optical color magnitude diagrams}

The construction of an optical CMD is not as straightforward as the
NIR CMD described above, due to the sensitivity and saturation limits of the available optical
images.
The optical photometry catalog of \citet{Baume03} is reported to become incomplete
at $V = 16$, i.e.,~it will obviously miss a large fraction of the
low-mass stars of NGC~3293.
The VST catalog from the VPHAS+ survey is considerably deeper, but misses
all stars brighter than $g \approx 13$, due to saturation effects.
Another problem is that these two catalogs use different photometric
systems, $UBVRI$ for \citet{Baume03} and SDSS $ugri$ for VPHAS+.
We  therefore consider the bright and the faint parts of the optical
CMD separately.

The upper plot in Fig.~\ref{cmd-opt.fig} shows the bright part of the CMD,
based on the \citet{Baume03} data.
All stars with $V \la 9$ (i.e.,~the most luminous early B-type stars)
are detected in X-rays. This matches the expectations since these very bright objects
are the most luminous early B-type stars, which are
expected to be strong X-ray emitters because they have quite strong
stellar winds  and their X-ray luminosity 
is correlated to their bolometric luminosity \citep[see][]{Gagne11}. 
At somewhat fainter magnitudes, in the 
range  $V \approx [9\!-\!13]$, only a small fraction of the
stars are detected in X-rays; this range
corresponds to stars with late B and
A spectral types, which are not expected to be intrinsic X-ray emitters.
A star-by-star discussion of the individual X-ray properties and 
the observed trends of X-ray emission as a function of spectral type and
stellar mass in the B- and A-type range 
is provided in Sec.~\ref{Bstar.sect}.
At fainter magnitudes, $V \ga 13$ (corresponding to stellar masses
$\la 3\,M_\odot$), the number of X-ray detected stars
 increases strongly,
as expected for a stellar mass function that rises towards lower masses.

\subsubsection{ Age of the low-mass population}

As can be seen in Fig.~\ref{cmd.fig} and Fig.~\ref{cmd-opt.fig},
most of the X-ray selected stars lie close to or to the right of the 8--10~Myr 
isochrones.
Taking the uncertainties of the photometry into account ($\sim 0.05$~mag),
the CMDs are consistent with the assumption that the majority of the X-ray 
selected low-mass stars have ages of $\sim 8-10$~Myr.
This result is in good agreement with the above-mentioned previous 
age estimates for NGC~3293, which
were based on CMD positions of the high-mass stars and the comparison with
post-main-sequence evolutionary tracks.
This consistency suggests 
a common age of about 10~Myr for the high- and  the low-mass stars
in NGC~3293.

\subsubsection{Size of the X-ray selected stellar population of NGC~3293}

Before we can quantify the population of X-ray detected young stars in NGC~3293,
we first have to define the area over which we assume X-ray detected stars to 
be cluster members.
As shown in Sect.~\ref{spatial-distr.sec}, the stellar cluster has a
radial extent of about 7~arcmin.
This value for the radius is also supported by the fact
that it just encloses all those 45 B-type stars for which \citet{Dufton06}
determined a stellar mass of $M \ge 5\,M_\odot$.
This circular region also contains the large majority of all X-ray sources
with infrared counterparts, supporting the chosen value.

We determine the cluster population from the NIR CMD of the X-ray detected stars, since --
as described above --
the optical VST data are not sensitive enough to detect all the
X-ray detected stars in NGC~3293.
In the $J$ versus $J-H$ diagram in Fig.~\ref{cmd.fig},
the number of X-ray detected stars with CMD positions that are
consistent with young stellar members of NGC~3293 (i.e.,~at most 0.05~mag 
[the $1\sigma$ uncertainty of the photometry] 
to the left of the 10~Myr isochrone,
or to the right of the 10~Myr isochrone, and above the reddening vector starting from the
location of 10~Myr old $0.1\,M_\odot$ stars) in the 7 arcmin radius region 
is 511.
We note that the inclusion of the \textit{Chandra} sources that have point-like counterparts
in the VISTA images that are missing from the VISTA catalog 
and photometry with $\leq 20\%$ uncertainties would raise that number by
31, i.e.,~to 542.

This number can now be compared to the
numbers of X-ray detected stars in the other clusters in the CNC,
as listed in Table~1 of  \citet{CCCP-Clusters}. Since these other clusters 
were also studied with \textit{Chandra} observations of almost identical 
sensitivity \citep[see][]{CCCP-intro},
a quantitative comparison is straightforward.
The three most populous clusters in the central Carina Nebula are Tr~14 with 
1378 X-ray detected stars,
Tr~15 with 481 X-ray detected stars, and
Tr~16 with 530 X-ray detected stars.
{This comparison shows that NGC~3293 is 
clearly one of the most populous clusters in the entire CNC.
It is less populous than Tr~14, but almost equal to 
Tr~16 and Tr~15, and  more populous than 
all the other known clusters in the CNC}. NGC~3293 is thus an important part
of the CNC (see discussion in Sect.~\ref{conclusions.sec}).

\subsubsection{Size of the low-mass population and the mass function of NGC~3293  \label{IMF_counts.sec}}

New information about the mass function of the cluster can be obtained
by comparing the
number of X-ray detected low-mass stars in a given mass range to
the known number of high-mass stars that were identified in
optical spectroscopic studies. This will also provide 
an important clarification about the previous claims of a significant deficit
of low-mass stars in NGC~3293 that were made in some optical 
photometric studies \citep{Slawson07,Delgado11}.

Our aim here is to determine the number of stars in the $[1-2]\,M_\odot$ mass range.
To this end, we count the number of X-ray detected
stars at positions in the color magnitude diagram that are consistent
with masses between $1\,M_\odot$ and $2\,M_\odot$ for an age of $8\!-\!10$~Myr;
this concerns all objects with positions in a
$\pm 0.05$~mag band around the 10~Myr isochrone line for $[1-2]\,M_\odot$
or at locations to the right of this isochrone shifted along the direction 
of the reddening vector (as indicated by the yellow-and-black lines in Fig.~\ref{cmd.fig}
and Fig.~\ref{cmd-opt.fig}).
The total number of such VISTA catalog stars in the 7 arcmin cluster region
is 179 in the $J$ versus $J-H$ diagram.
In the optical $g$ versus $g-r$ diagram, the corresponding number of 
VST catalog stars
is 168 in the same 7 arcmin cluster region.

In order to estimate the underlying stellar population, this
number of X-ray detected stars has to be
 corrected for the finite X-ray detection completeness,
which is also a function of the location of the stars in the ACIS-I field
(due to the variation of the X-ray sensitivity with off-axis angle).
In order to estimate what fraction of the stars in a certain
mass range can be detected  at a specific local detection limit,
we use here the 
X-ray luminosity functions of young stars
derived in the context of the \textit{Chandra} Orion Ultradeep Project (COUP)
\citep{PF05}.
In the central part of our \textit{Chandra} image, the weakest detected
X-ray sources  
have X-ray luminosities of $\log (L_{\rm X} [{\rm erg/s}]) \simeq 29.6$, 
if we assume a thermal plasma with $kT = 1$~keV 
and an absorbing column density of $N_{\rm H} = 2 \times 10^{21}\,{\rm cm}^{-2}$
(corresponding to $A_V \simeq 1$~mag).
A comparison to the X-ray luminosity function of solar-mass $[0.9\,M_\odot \le M \le 1.2\,M_\odot]$ 
stars with ages of a few Myr in \citet{PF05}
shows that $\simeq 90\%$ of the stars will be above this limit.
With an age of $\approx 8\!-\!10$~Myr, NGC~3293 is somewhat older, 
and the X-ray luminosities of the stars are thus expected to be 
slightly lower;
on the other hand, stars slightly more massive than one solar mass 
will have somewhat higher X-ray luminosities.
Therefore, assuming a completeness factor of $\simeq 90\%$ is a 
reasonable choice for coronally active young stars in the center of NGC~3293 
in the mass range $\approx [1\!-\!2]\,M_\odot$.

Since the cluster has a spatial extent  of several arcminutes, we 
have to take the variation of the X-ray detection limit as a function
of off-axis angle into account.
For this, we use the results derived in the detailed study of \citet{CCCP-catalog}
and summarized in their Table 8, where the variation of the X-ray luminosity limit is determined
for different ranges of the off-axis angle. As listed there,
the X-ray luminosity limit in the off-axis angle range $[3.8\!-\!6.3]$~arcmin
is 0.5 dex higher than in the central 3.8~arcmin; in the $[6.3\!-\!7.5]$~arcmin off-axis angle range
it is 0.6 dex higher than in the central 3.8~arcmin.
From these numbers, we find  X-ray luminosity function completeness factors
of $\simeq 60\%$ for stars in the off-axis angle range $[3.8\!-\!6.3]$~arcmin and
 $\simeq 59\%$ for stars in the off-axis angle range $[6.3\!-\!7.5]$~arcmin.

Counting the X-ray detected stars in the 7 arcmin cluster region
that have positions in the $J$ versus $J-H$ diagram
consistent with 
 masses  between $1\,M_\odot$ and $2\,M_\odot$ for an age of $8\!-\!10$~Myr
according to the PARSEC stellar models, we
find the following numbers for the different off-axis angle ranges:
  in the central part ($\theta = [0-3.8]$~arcmin) we find 104 stars,
in the $\theta = [3.8-6.3]$~arcmin range we find 70 stars,
and in the $\theta = [6.3-7.5]$~arcmin range we find 5 stars.
Dividing these numbers by the corresponding X-ray luminosity function completeness factors
of 0.90, 0.60, and 0.59, yields a total extrapolated number of 241 stars 
as our final estimate for the total number of stars in the $[1-2]\,M_\odot$ 
mass range within the 7 arcmin cluster region. 
Considering the uncertainties in the determination of the
X-ray luminosity function completeness factors, we use 
$N[1\!-\!2\,M_\odot] \simeq 241 \pm 25$ stars 
as the final result of this analysis.

We note that the true number is somewhat higher, since we omit here the 61 VISTA point source counterparts
to \textit{Chandra} sources that are lacking photometry in the VISTA catalog.
If we  include  the  stars from this group of counterparts that have  $<20\%$ photometric uncertainty into
the counting,
there will ten additional stars in the central off-axis bin, and the
extrapolated total number estimate will be $N[1\!-\!2\,M_\odot] \simeq 252$
instead of 241.

As discussed in Sect.~\ref{reliability.sec}, we have to 
expect between about five and nine random false matches 
to sufficiently bright VISTA catalog sources that might be
located in our counting region for the $[1\!-\!2]\,M_\odot$ stars in the 
CMD. These estimated false matches have to be subtracted
from our estimate of the stellar population.
However, we should also  note that six of the ten above-mentioned  VISTA sources 
missing from the VISTA catalog that correspond to X-ray detected stars
in the $[1\!-\!2]\,M_\odot$ range, have an optical counterpart in the 
VST VPHAS-DR2 point source catalog.
At least these six additional VISTA stars should thus certainly  
be included into the counts. 
The difference between the five to nine objects that should be 
subtracted and the six objects that should be added is small
compared to the  above given uncertainty range $\pm 25$.
Therefore, we conclude that our result of 
$N[1\!-\!2\,M_\odot] \simeq 241 \pm 25$
is robust.

This number can now be compared to the
size of the high-mass population in the same $R = 7\arcmin$ region.
In addition to the above-mentioned 45 B-type stars with listed masses between 
$5\,M_\odot$ and $40\,M_\odot$ \citep{Dufton06}, the red supergiant V361~Car is
also located in this region.
The total number of high-mass stars in this cluster region
is thus 46.
Using the numerical representation of the
canonical field star IMF from \citet{Kroupa02} and
the number of 46 stars in the $[5-40]\,M_\odot$ range, the
expected number of $[1-2]\,M_\odot$ stars is $N_{\rm IMF\,exp} = 237$.
The X-ray completeness-corrected number of 
$N[1\!-\!2\,M_\odot] \simeq 241 \pm 25$ such stars 
($253 \pm 25$, if the stars missing in the VISTA catalog are included)
derived above
is in very good agreement with this expectation value.

This good agreement suggests that the size of the solar-mass 
population of NGC~3293 
is  consistent with the expectations from the normal field star IMF.
This result refutes earlier claims for a strong deficit of 
stars with mass below  $M \le 2-3\,M_\odot$.
This highlights the difficulties resulting in purely photometric determinations 
of cluster populations, especially in regions with a very strong galactic background
such as in NGC~3293.

We cannot directly determine the IMF in the subsolar and very low-mass 
range with our data
since the X-ray detected sample is very  incomplete at such low stellar
masses.
Assuming that the IMF of NGC~3293 follows the field IMF down to $0.1\,M_\odot$,
a numerical extrapolation (based again on the 46 high-mass stars) of the field star IMF
suggests 3230 stars in the $[0.1\!-\!1]\,M_\odot$ range.
In the same way, an estimate of the total stellar population of
3625 stars in the $[0.1\!-\!100]\,M_\odot$ range can be computed.

The statistical expectation value of the number of very high-mass stars (above $40\,M_\odot$) is 2.3. The central 68\% Poisson range
is $[ 1\!-\!4]$ such stars, which 
should have already exploded as supernovae.

\section{X-ray properties of the  B-type stars \label{Bstar.sect}}

%                                              
%-----------------------------------------------------------
\begin{figure}
\centering
\includegraphics[bb=5 15 432 338,width=8.5cm]{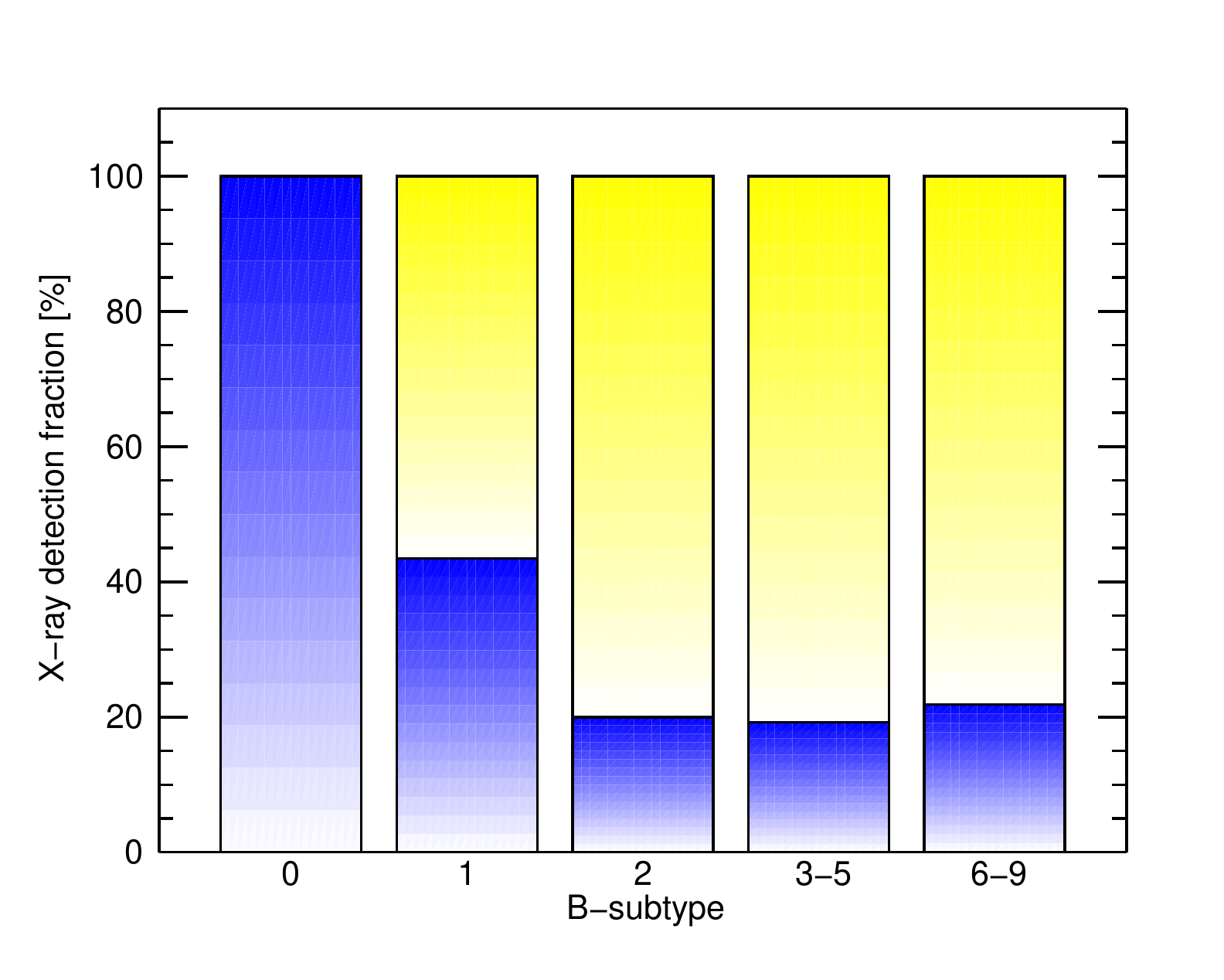}
\caption{Fraction of X-ray detections (blue shaded histogram) as a function of
spectral type.
}
\label{Bstar-Xdet-stat.fig}
   \end{figure}
%
%______________________________________________________________

The large number of B-type stars in NGC~3293
provides a good opportunity to investigate
the X-ray properties in a co-eval and homogeneous population of 
B-type stars, covering the full spectral range from B0 to B9.

It is well established that most of the hottest B-type stars (spectral types 
B0 to $\approx$~B2) are rather strong X-ray sources; their
X-ray emission is thought to be related to their strong stellar winds,
similar to the case of O-type stars \citep[e.g.,][]{Zhekov07,Cohen14}.
Most B-type stars with spectral types later than $\approx$~B2
remain undetected in X-ray observations; this is in good agreement
with the theoretical expectations, since these stars only have  relatively
weak winds that are  incapable of producing strong X-ray emission,
and at the same time the B-type stars (and also the A-type stars)
have no outer convection zones, and thus no magnetic dynamo action
is expected, which is the prerequisite for the X-ray emission
due to coronal magnetic activity (as in the late-type stars).

Nevertheless, in many young clusters, a significant
fraction of late B-type stars
shows detectable amounts of X-ray emission \citep[e.g.,][]{Stelzer05,Gagne11}.
An often invoked explanation is the presence of unresolved late-type
companions as the true source of the X-rays. The X-ray properties of
these late B-type stars will then contain information about the
multiplicity and the nature of their low-mass companion stars.

\subsection{X-ray detection fraction as a function of spectral type}

In the full sample of 97 B-type stars in the ACIS-I field, 24 are detected as
X-ray sources. The detection fraction, however, is a strong function of spectral type,
as shown in Fig.~\ref{Bstar-Xdet-stat.fig}.
While all (three) B0 stars
are detected, the fraction drops to $10/23 \simeq 43\%$ for the B1 stars, 
and then to $\approx 20\%$ for later B spectral types.

If we assume that for all of these X-ray detected later B-type stars
the X-ray emission originates from an unresolved late-type companion,
information about the multiplicity and the pairing-statistics of these
B-type stars can be inferred.
 Since X-ray luminosity scales with stellar mass, these companions
cannot have stellar masses that are too low; a reasonable guess is that these
companions should be stars with $M \ga 1\,M_\odot$ in order to 
produce X-ray luminosities above our detection limits.

If we assume that every B-type star has one companion (i.e.,~a multiplicity of 100\%), 
and further assume that the companion masses are randomly sampled from the conventional field-star 
IMF, it can easily be calculated that the expected fraction
of B-type stars with companions of mass $\ge 1\,M_\odot$ would be\footnote{According to the 
\citet{Kroupa02} IMF, 
the ratio of stars in the \mbox{$[1-2]\,M_\odot$} range to those in the $[0.1-2]\,M_\odot$ range
is 0.068.}
$\approx 6.8\%$.
This value is lower than the observed X-ray detection rate of $\approx 20\%$
and thus implies that either
 these B-type stars have more than one companion on average or
that the companion masses are not established by random sampling,
but are biased towards more massive stars.
This supports independent observational results 
which also suggest a very high multiplicity of intermediate- to high-mass
stars \citep[see, e.g.,][]{Preibisch99,Duchene13} and mass ratios that are
higher than expected from random pairing. Our result also agrees with previous findings
based on X-ray observations of other young clusters \citep[see, e.g.,][]{Evans11}.

\subsection{X-ray properties of the individual B-type stars}

For a more quantitative analysis, we
determined the X-ray properties of the individual B-type stars.
For those four B-type stars with a sufficient number of X-ray counts,
spectral fitting was performed with XSPEC.
In the other cases, we used the conversion factor from counts
to unabsorbed flux as determined by \texttt{srcflux},
 employing the
optically derived individual extinction of each star to fix the 
hydrogen column density, and assuming a plasma temperature of 
$kT = 0.5$~keV for these stars.
The \texttt{srcflux} tool was also used for those B-stars without
an X-ray counterpart in our source list, in order to compute 90\% upper
limits to their count rates and X-ray luminosities.

In the following, we briefly discuss the individual X-ray properties of some of the
X-ray detected B-type stars.
The full set of information about the X-ray luminosities (or the corresponding upper limits) 
for the 
B-type stars is contained in the electronic Table 3, available at the CDS.
\medskip

\noindent{\bfseries HD~91969:}\\
The optically brightest ($V = 6.51$) 
cluster member 
is the B0~Ib star HD~91969, for which a stellar mass of $M \approx 40\,M_\odot$
has been estimated \citep{Dufton06}. 
HD~91969 provides  a perfect match to the \textit{Chandra} source 542, 
which yielded 164.4 net counts and shows a
rather soft spectrum with a median photon energy of 1.0~keV.
The XSPEC fit with a thermal plasma model and
the $N_{\rm H}$ parameter fixed to the value 
$0.136 \times 10^{22}\,\mathrm{cm}^{-2}$ (as determined from the
optical color excess)
yields a plasma temperature of $T \approx 5.7 \pm 1.1$~MK. 
This is within the typical range of
plasma temperatures of early B-type stars, for which the X-ray emission
is assumed to originate from shocks in the fast stellar wind
\citep{Stelzer05}.
Since the fit is of mediocre quality ($\chi^2/\nu = 2.92$), we also
tried models with two plasma components;  however, they yielded no
improvement to the fit quality.
The  X-ray luminosity of the best-fit model is
$L_{\rm X} = 2.78 \times 10^{31}$~erg/sec and yields a
fractional X-ray luminosity of $L_{\rm X} / L_{\rm bol} = 3.0 \times 10^{-8}$,
close to the typical ratios found for early B-type stars.

We note that two further (but much weaker)
X-ray sources were detected very close to HD~91969:
 source 537, located $1.4''$ to the west of HD~91969 
with 4.2 net counts, and source 548, 
located $3.0''$ to the northeast of HD~91969
with 6.9 net counts.
In the available optical and NIR images, no indication of the
presence of stars is seen at these locations because  the source positions
are within the very bright (or even saturated) parts of the PSF of the
extremely bright star HD~91969.
It therefore remains unclear, whether these two X-ray sources might be 
(presumably late-type) companions to HD~91969. 
%--------------------------------------------------------------------
\medskip

\noindent{\bfseries HD~91943:}\\
The optically second brightest ($V = 6.69$) star, HD~91943, has a 
spectral type B0.7~Ib and an estimated stellar mass of 
$M \approx 30\,M_\odot$.
This star provides a perfect match to the X-ray source 
418, which yielded 38.8 net counts with
a median energy of 1.1~keV.
The fit to the X-ray spectrum yielded a plasma temperature of
 $T = 8.5 \pm 4.6$~MK and an X-ray luminosity of 
$L_{\rm X} = 4.1 \times 10^{30}$~erg/sec.
This corresponds to a fractional X-ray luminosity of 
$L_{\rm X} / L_{\rm bol} \approx 7 \times 10^{-9}$.

No further X-ray source is detected within $8\arcsec$ of HD~91943.
\medskip

\noindent{\bfseries CPD~$\mathbf{-57^{\circ}3524A}$:}\\
This B0.5~IIIn star is a good match to \textit{Chandra} source 704;
with 20.8 net counts, the X-ray source
is too weak for a spectral fitting analysis, but the relatively low
median energy of 1.3~keV is in the typical range for B-type stars
\citep[see Fig.~4 in][]{Gagne11}.

\medskip

%-------------------------------------------------------------------

\noindent{\bfseries CPD~$\mathbf{-57^{\circ}3526}$B:}\\
This B1 III star  
 is a very good match to \textit{Chandra} source 710,
which has 4.6 net count and a median energy of 
$1.3$~keV. Although the X-ray detection is unquestionable, the
X-ray properties remain somewhat unclear because  the
source detection revealed another, similarly strong source with number 709 (3.5 net counts,
 $E_{\rm med} = 0.8$~keV) at an angular distance of just $0.68\arcsec$ from
source 710.

Unfortunately, in all the available optical and infrared images of this 
very bright star ($V = 8.25$,  $H = 8.03$) the inner few square arcseconds
of the PSF are completely saturated; it is therefore not possible to
check whether the second X-ray source is a companion to the
B1 star.

%-------------------------------------------------------------------

\medskip

\noindent{\bfseries CD~$\mathbf{-57^{\circ}3348}$:}\\
The B1~III star CD~$-57^{\circ}3348$ (alias CPD~$-57^{\circ}3506$A)
is a very good match to \textit{Chandra} source 490, 
which yielded 42.8 net counts with a median energy of 1.0~keV.
The fit to the X-ray spectrum yielded a plasma temperature of
 $T = 7.9 \pm 4.5$~MK and an X-ray luminosity of
$L_{\rm X} = 5.4 \times 10^{30}$~erg/sec.
This corresponds to $L_{\rm X} / L_{\rm bol} \approx 2.3 \times 10^{-8}$.

We note that the VISTA images show another fainter star $3\arcsec$ to the north of
this B1~III star, which is also detected as an X-ray source.
Since these $3\arcsec$ correspond to a physical distance of about 7000~AU,
it remains unclear whether this fainter star might be a companion to the
B1~III star, or just a random projection effect.
%-------------------------------------------------------------------
\medskip

\noindent{\bfseries CPD~$\mathbf{-57^{\circ}3523}$:}\\
This B1~III star is a very good match to \textit{Chandra} source 697. 
With 14.8 net counts the X-ray source is too weak for spectral fitting,
but the median energy of 0.9~keV is in the typical range for early
B-type stars.
\medskip

\noindent{\bfseries V405~Car:}\\
The B1 V star V405~Car (alias  CPD~$-57^{\circ}3507$)
 is a very good match to \textit{Chandra} source  523,
which has 3.8 net counts and $E_{\rm med} = 2.7$~keV; 
this is unusually hard for a B star
and thus indicates that a low-mass companion contributes to the observed
Xray emission.
%-------------------------------------------------------------------
\medskip

\noindent{\bfseries CPD~$\mathbf{-57^{\circ}3521}$:}\\
The B1~III star  CPD~$-57^{\circ}3521$ is a very good match to \textit{Chandra} source 679,
which has 14.7 net counts and $E_{\rm med} = 1.4$~keV.

The
X-ray analysis revealed another, weaker (2.8 net counts) but somewhat harder
 ($E_{\rm med} = 2.0$~keV) X-ray source (680) 
at an angular distance of just $0.82\arcsec$ from
source 679.
Unfortunately, the reality of this tentative (late-type?) companion
cannot be checked since in
all available optical and infrared images of this
very bright star ($V = 8.14$,  $H = 7.87$) the inner few square arcseconds
of the PSF are completely saturated. 
%-------------------------------------------------------------------
\medskip

\noindent{\bfseries NGC~3293 ESL~87:}\\
The B5 star NGC~3293 ESL~87 
is a good match to \textit{Chandra} source 47, 
which has 25.9  net counts and $E_{\rm med} = 1.6$~keV.
The XSPEC fit to the X-ray spectrum (with $N_{\rm H}$ fixed at
$0.1\times 10^{22}\,{\rm cm}^{-2}$) yields a plasma temperature of 
$kT = 2.1 \pm 1.5$~keV and an X-ray luminosity of $4.6 \times 10^{30}$~erg/sec.
These values would be very unusual for a B5 star and thus point towards 
the presence of a late-type companion.

%                                              
%-----------------------------------------------------------
   \begin{figure}
   \centering
\includegraphics[width=8.85cm]{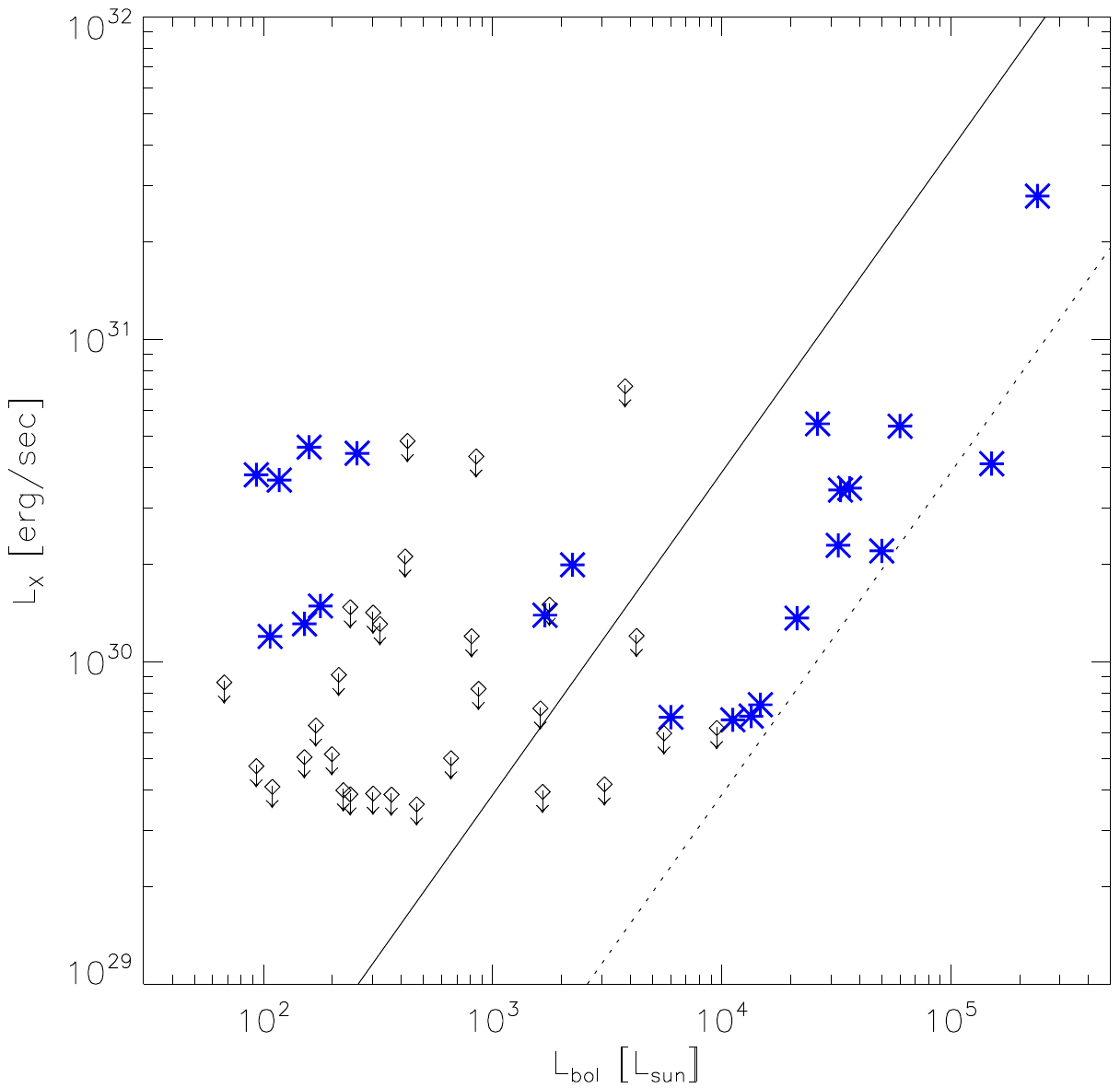}
      \caption{X-ray versus bolometric luminosity of the known B-type
stars in NGC~3293. The blue asterisks show the X-ray detected stars.
Upper limits for the non-detected B-type stars are marked
with the small diamonds with downward-pointing arrows. The solid
line marks the relation   $L_{\rm X} / L_{\rm bol} = 10^{-7}$, the
dotted line $L_{\rm X} / L_{\rm bol} = 10^{-8}$.
}
         \label{lxlbol-Bstars.fig}
   \end{figure}
%
%______________________________________________________________

\bigskip

\noindent{\bfseries Other B-type stars in NGC~3293:}\\
Most of the other B-type stars in the ACIS-I field
are not detected as X-ray sources.
As described above, we use the \texttt{srcflux} tool to
determine 90\% upper limits to the X-ray luminosities
of the undetected B-type stars. 
This yielded useful upper limits for 31 of the undetected B-type stars;
for the remaining targets, \texttt{srcflux} could not determine
upper limits because  they are located very close to a chip edge on the detector.
The upper limit values determined in this way reflect the
local X-ray detection limit at the target position.
The three B-type stars with particularly high
($\ge 2\times 10^{30}$~erg/s) upper limits (see Fig.~\ref{lxlbol-Bstars.fig})
are located very close to the edge of the ACIS-I detector where the 
sensitivity is considerably lower than on-axis.

\subsection{X-ray and bolometric luminosity}

In Fig.~\ref{lxlbol-Bstars.fig} we plot the X-ray luminosities
of the detected B-type stars (and the upper limits to the 
X-ray luminosities of the undetected  B-type stars) against
the bolometric luminosities.

Most of the very luminous objects ($L_{\mathrm{bol}} \ga 10^4\,L_\odot$) 
show fractional X-ray luminosities between
$L_{\rm X} / L_{\rm bol} = 10^{-7}$ and $L_{\rm X} / L_{\rm bol} \approx 
10^{-8}$. This is the typical range found for early B-type stars
in other young clusters \citep[e.g.,][]{Stelzer05,Gagne11} and is
consistent with the wind-shock model for the origin of the X-ray emission.

The X-ray detected later B-type stars show considerably higher $L_{\rm X} / L_{\rm bol}$
ratios,
which would be very hard to explain by the wind-shock mechanism thought
to be at work in the early B-type stars. Their X-ray luminosities
are, however,  consistent with those of young late-type stars,
again strongly suggesting that the X-ray emission actually originates
from unresolved late-type companion stars.

\subsection{ Red supergiant star V361 Car (M1.5Iab-Ib)}

In order to describe all known massive stars in NGC~3293, we note
that the red supergiant star V361~Car (M1.5Iab-Ib) (which is the most
massive star in the cluster) is not detected as an X-ray source.

However, the  X-ray source 449
is found just 2.8 arcsec southeast of V361~Car.
With 4.9 net counts and  a relatively soft spectrum ($E_{\rm med} = 1.2$~keV),
this might be either a late-type companion, or  an early B-star companion 
to V361~Car.
Owing to the extreme brightness of the supergiant ($V = 7.19$, $H = 2.6$),
all available optical and infrared images 
are completely saturated at the position of this possible companion.

\section{Summary and conclusions \label{conclusions.sec}}

We have performed the first deep X-ray observation of the young cluster
NGC~3293 at the northwestern edge of the Carina Nebula Complex.
This \textit{Chandra} observation complements the similarly sensitive X-ray
survey of the central region of the CNC in the context of the CCCP \citep{CCCP-intro}
and our recent observation of the NGC 3324 region (between the central parts
of the CNC and NGC~3293), and it
completes the X-ray investigation of all the star clusters
in the CNC.
The present study is the first where NGC~3293 is investigated
in the same way 
as all other clusters in the CNC (i.e.,~by a combination of a deep X-ray and infrared data).
This finally allows us to put the derived properties of NGC~3292 into a larger context and
to consider it in the same way as the other parts of the CNC.

Our analysis of the \textit{Chandra} X-ray observation of NGC~3293 clearly shows
that the cluster hosts a large population of low-mass stars in the
$\sim\![2\, -\, 0.5]\,M_\odot$ mass range.
The number of X-ray detected low-mass stars closely agrees  with the 
expectations based on the number of spectroscopically identified high-mass stars
and the assumption of a field star initial mass function.
There is thus {no} indication of a deficit of low-mass $(M \le 2.5\,M_\odot)$ stars in this cluster.
These results suggest a total population of $\approx 3600$~stars for NGC~3293.

We find that NGC~3293 is one of the most populous clusters
in the entire Carina Nebula Complex. NGC~3293 is older than the other  well-investigated
clusters in the CNC. Extrapolating the cluster's mass function suggests that several
supernova explosions  have occurred in NGC~3293 during the last few Myr
\citep[see also][]{Voss12}.
This suggests that NGC~3293 has most likely
played an important role during the
formation and early evolution of the CNC.

With a spatial extent of $\sim 100$~pc, a total cloud mass of
$\sim 10^6\,M_\odot$, and more than 100\,000 young stars,
 the CNC is one of the largest
star forming complexes in our galaxy. It also shows a high
diversity in the structure of the clouds and in the spatial
configurations of the stellar populations.
Most of the
 stars in the inner parts of the CNC are located
in one of more than ten individual stellar clusters, which have ages 
ranging from $\la 1$~Myr  up to
$\sim 5$~Myr. 
In addition to this clustered stellar population, there is also
an unclustered, widely
distributed population \citep[see][]{CCCP-Clusters} of young stars with
ages ranging from $< 1$~Myr  to $\sim 6-8$~Myr \citep{CCCP-HAWKI}.
How this complicated and diverse spatial and temporal configuration
has formed and evolved is still unclear.

\begin{figure}
\centering
\includegraphics[bb=55 190 516 795,width=6.5cm]{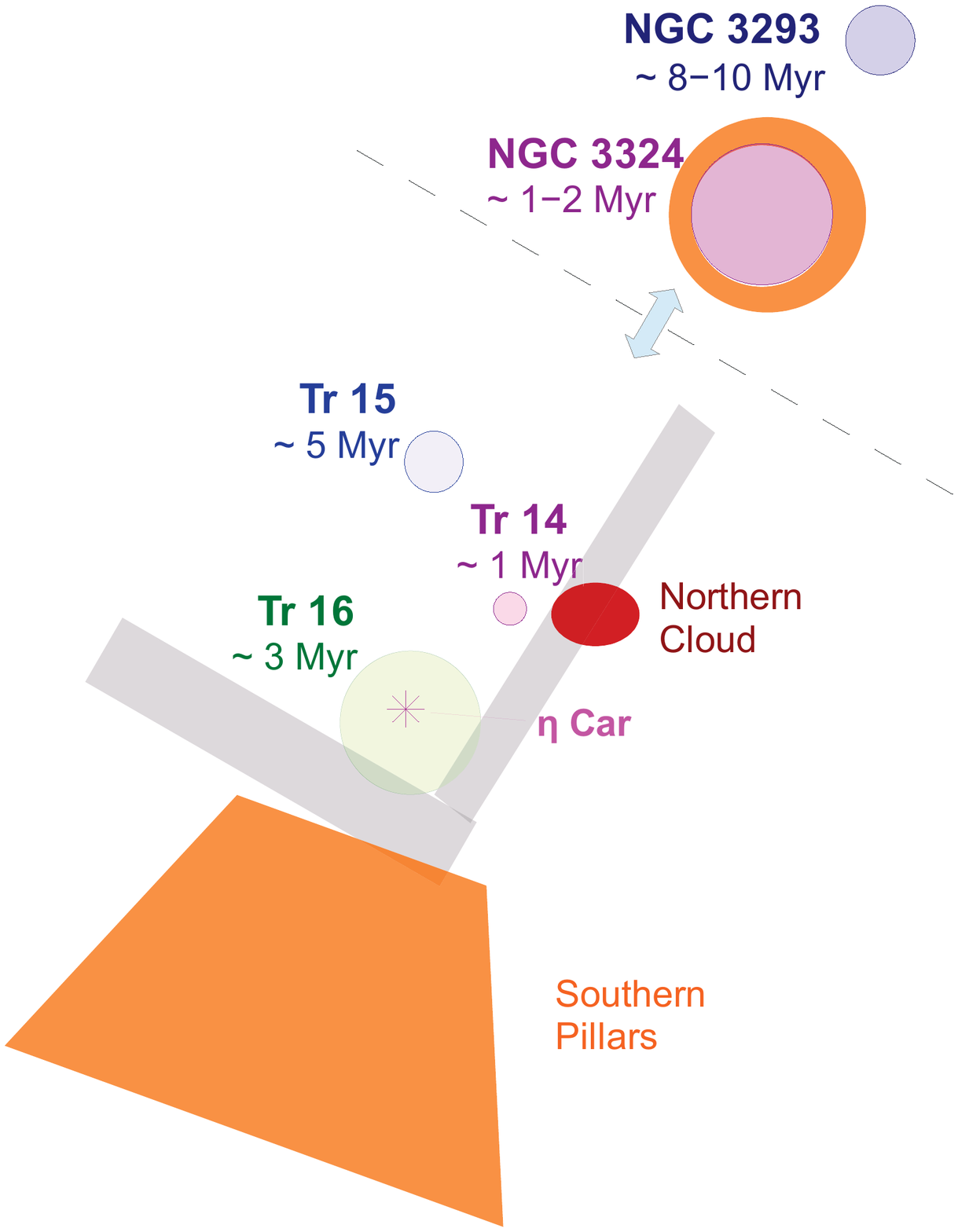}
\caption{Sketch illustrating the locations and orientations of the main clusters
and clouds in the Carina Nebula Complex. The circles indicate the locations and
the approximate sizes of the individual stellar clusters. The orange and red structures
indicate the approximate extent of the densest clouds, and the gray rectangles
mark the \textsf{V}-shaped lane of dark clouds intersecting the central parts of the
Carina Nebula.
As indicated by the dashed line and the double arrow, the distance of
NGC~3324 and NGC~3293 from the center of the Carina Nebula is actually somewhat greater
than in this sketch.
\label{CNC-sketch.fig}}
\end{figure}

In Fig.~\ref{CNC-sketch.fig} we show a sketch illustrating the spatial configuration
of the most significant clusters and clouds in the CNC.
 NGC~3293 is the oldest of the large clusters in the CNC,
and it is located at the northwestern edge of the complex.
Most of the current star formation activity observed today in the CNC is happening
in the southeastern part of the CNC, in the Southern Pillars region, i.e.,~at the
opposite end of the complex. However,
this does {not} reflect a systematic northwest to southeast spatial age gradient since
there are several sites of active star formation in the northern part of the complex.
The most prominent of these is the large shell of
dense dust clouds around the cluster NGC~3324
\citep[see][]{Ohlendorf13}, which is located
just southeast of NGC~3293.
Another example of very recent and ongoing star formation activity in the northern
area is the embedded stellar cluster associated with the massive dense cloud clump G286.21+0.17
\citep[see][]{Ohlendorf13},
which is located north of the dust shell around NGC~3324.

The cluster ages also show {no} systematic trend with spatial position in the complex. 
For example,
the very young  ($\sim\!1\!-\!2$~Myr) cluster NGC~3324 is located between the considerably older clusters
NGC~3293 ($\sim\!8\!-\!10$~Myr) to the north  and the $\sim 5$~Myr old cluster Tr~15 south of it.

Instead of a systematic progression of star formation in one direction, it appears that the
star formation activity in the CNC was ``wandering around'' in various directions that changed
with time. The best explanation of this seems to be that
the progression of star formation in the CNC was predominantly triggered by the
feedback of the numerous high-mass stars that were born at different locations in the complex over
the last $\sim 10$~Myr.

Today, we witness the effect of this feedback and the corresponding driving of
star formation everywhere in the surroundings of the currently existing O-type stars.
For example, the  $\sim 3$~Myr  high-mass stars in the cluster Tr~16 are triggering star formation
in the Southern Pillars \citep[see][]{Smith10b,Povich11}.
The very young ($\sim 1$~Myr) high-mass stars in the cluster
Tr~14 have recently started to exert a strong influence on the very dense and
massive cloud 
 \citep[known as the Northern Cloud; see][]{Brooks03,Kramer08,CNC-Laboca,HAWKI-survey} to the west 
of the stellar cluster;
in the near future, the ongoing compression
of this cloud is likely to initiate cloud collapse and trigger star formation.
The young ($\sim\!1\!-\!2$~Myr) O-type stars in the cluster NGC~3324 \citep{Preibisch14}
have swept up a huge dust bubble around the Gum~31 HII region and are
starting to trigger star formation in and around this bubble  \citep{Ohlendorf13}.

 About 8~Myr ago, the (then very young) cluster NGC~3293 probably contained
several O-type stars. These high-mass stars must have exerted a  similar influence on the
clouds that were present in the cluster's surroundings at that time.
It appears likely that this stellar
feedback initiated the sequence of local cloud collapse events in the original,
huge proto-Carina Nebula cloud, starting probably at the location where we now find the
second oldest cluster Tr~15. Given the age difference of $\sim\!3\!-\!5$~Myr between NGC~3293
and Tr~15, it seems possible that supernova explosions of the O-type stars may have
played an important role in the timing of the propagating star formation.

Today, i.e.,~about 8--10 Myr later, star formation is still
going on in some locations of the complex. Although a large fraction
of the original $\sim 10^6\,M_\odot$ gas
has already been heated and transformed to lower densities by the
effects of the radiation, winds, and supernovae of several generations of 
massive stars \citep{Preibisch12},
there is still a large reservoir ($\ga 20\,000\,M_\odot$) of dense clouds 
\citep[see][]{CNC-Laboca}
available for future star formation over the next millions of years.

These considerations show that
a good characterization of the ages and stellar populations of clusters like
NGC~3293 is a key factor for 
understanding the intricate spatio-temporal progression
of star formation in huge cloud complexes like the CNC.

%\dataset [ADS/Sa.CXO\#obs/16648] {Chandra ObsId 16648}

\begin{acknowledgements}
 We gratefully acknowledge funding for this project by the German
\emph{Deut\-sche For\-schungs\-ge\-mein\-schaft, DFG\/} project
number PR~569/9-1. 
Additional support came from funds from the Munich
Cluster of Excellence: ``Origin and Structure of the Universe''.
Townsley and Broos acknowledge support from the \textit{Chandra X-ray Observatory} 
general observer grant GO5-16003X and from the Penn State ACIS Instrument Team Contract 
SV4-74018, issued by the \textit{Chandra} X-ray Center (CXC), which is operated by the 
Smithsonian Astrophysical Observatory for and on behalf of NASA under contract NAS8-03060. 
 This research used software provided by the CXC in the application package \textit{CIAO}, 
and \textit{SAOImage DS9} software developed by the Smithsonian Astrophysical Observatory.
The VISTA infrared data used in this work are
based on observations made with ESO Telescopes at the La Silla Paranal Observatory 
under programme ID 088.C-0117.
The analysis used data products from observations made with ESO Telescopes 
at the La Silla Paranal Observatory under program ID 177.D-3023, 
as part of the VST Photometric H$\alpha$ Survey of the Southern Galactic 
Plane and Bulge (VPHAS+, www.vphas.eu). 
We acknowledge the assistance of the LMU physics students
S.~Gra{\ss}l, J.~Diehl, and L.~Furtak
in some steps of the preliminary data analysis.
This research has made use of the SIMBAD database and the VizieR catalog services
operated at Strasbourg astronomical Data Center (CDS).
\end{acknowledgements}

\bibliographystyle{aa}
\bibliography{30874ref}

\begin{appendix}

\section{Cloud structure and extinction in and around NGC~3293\label{clouds-extinction.sec}}

\begin{figure*}
\parbox[t]{18cm}{\includegraphics[width=8.85cm]{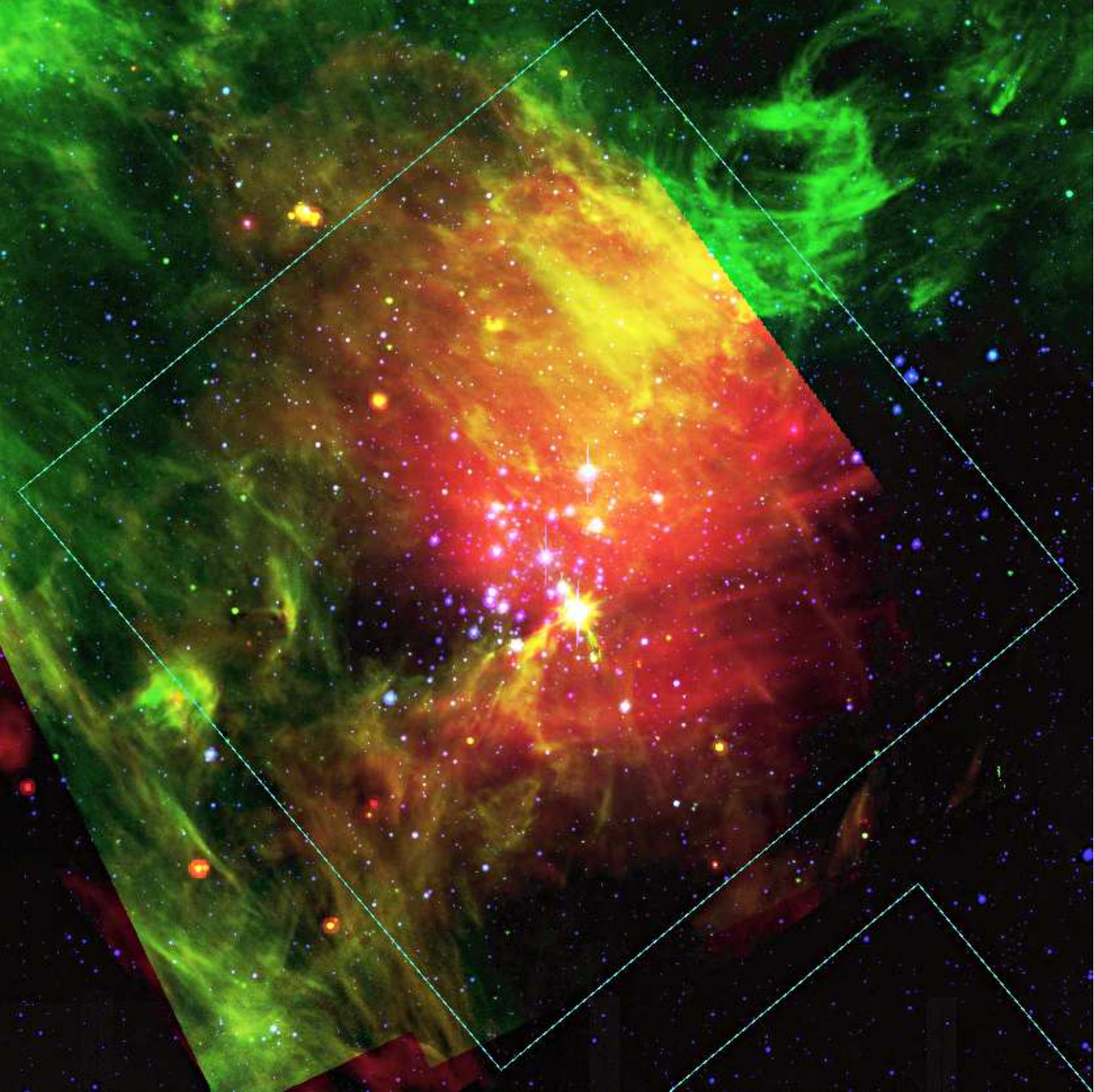}\hspace{3mm}
\includegraphics[width=8.85cm]{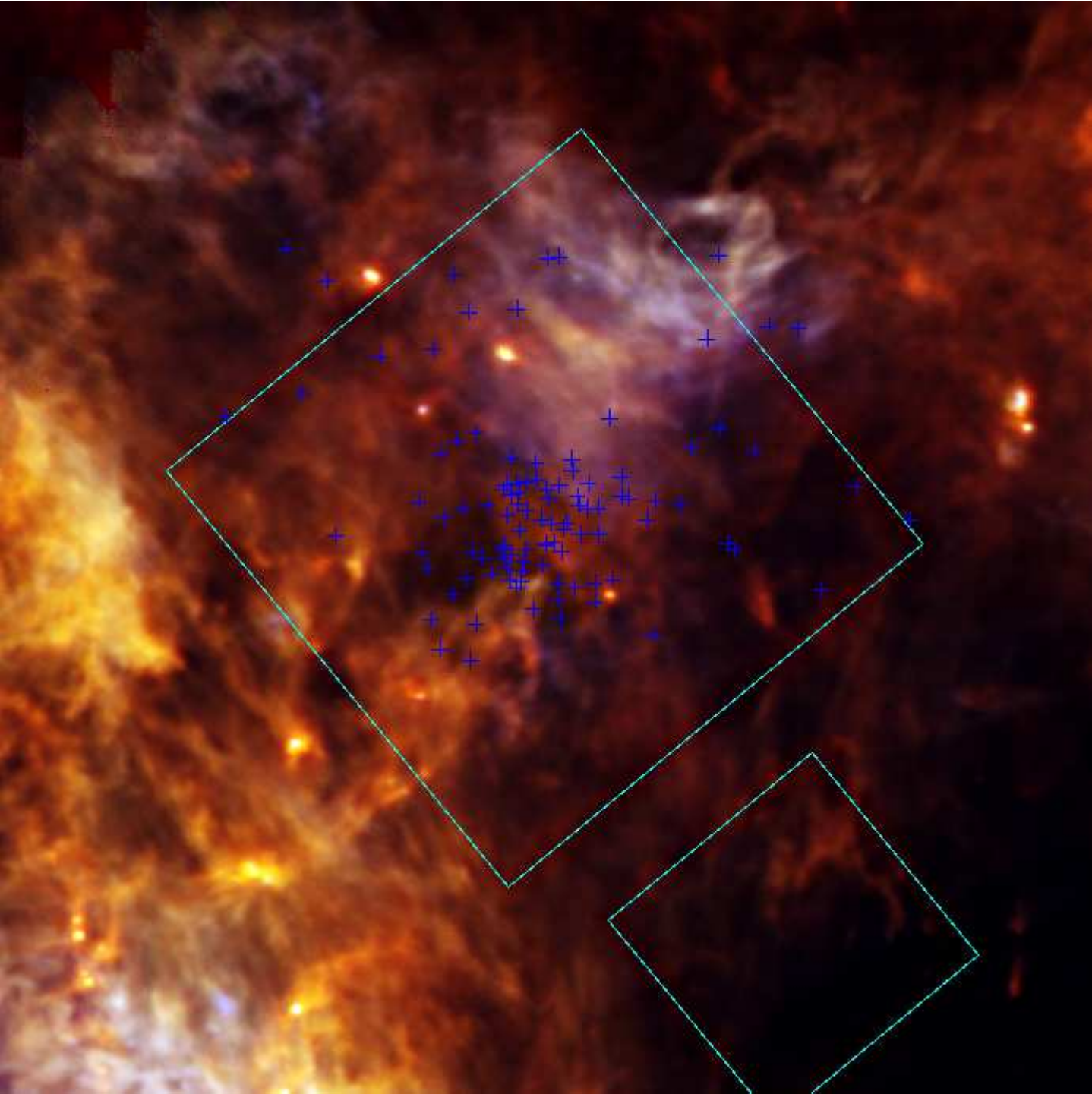}}
\caption{Left: Three-color composite image of the NGC~3293 area,
showing the
VST $r$-band image (from the VPHAS+ survey) in the blue color channel,
the \textit{Spitzer} IRAC $8\,\mu$m image (retrieved from the \textit{Spitzer}
 archive) in green,
and the \textit{Spitzer} MIPS $24\,\mu$m image in red.
The \textit{Spitzer} images cover only parts of the full
area shown in this image.
North is up and east to the left.
The cyan rectangle marks the $17\arcmin \times 17\arcmin$
area covered by our \textit{Chandra}
observation with the ACIS-I array, the lower cyan rectangle (extending beyond
the edge of the image) marks the area covered by the single active ACIS-S
detector.\newline
Right: Three-color composite image showing the \textit{Herschel}
PACS $70\,\mu$m image in the blue color channel, the
PACS $160\,\mu$m image in green,  and the SPIRE $250\,\mu$m image in red;
these \textit{Herschel} maps were obtained as part of the Hi-Gal survey
and were retrieved from the archive.
The B-type stars listed in \citet{Evans05} are marked by blue crosses.
The cyan rectangles mark again the \textit{Chandra} ACIS field of view.
\label{NGC3293-clouds.fig}}
\end{figure*}

NGC~3293 is located at the northwestern edge of the cloud complex
associated with the Carina Nebula. 
The inner parts of this cloud complex host some very dense
clouds, which cause very strong extinction of $A_V \ga 10$~mag
\citep[see cloud column density map in][]{Preibisch12} at some locations.
The properties and the highly inhomogeneous spatial
structure of these dense clouds have been revealed by
APEX sub-mm observations \citep{CNC-Laboca}. 

In the NGC~3293 area, the APEX / LABOCA maps from the ATLASGAL survey 
\citep{Schuller09}
show no significant sub-mm emission. Although this implies that no very
dense clouds are present, \textit{Spitzer} mid-infrared and \textit{Herschel}
far-infrared maps (see Fig.~\ref{NGC3293-clouds.fig}) 
nevertheless show significant emission from moderate-density clouds in and
around NGC~3293.
The diffuse $24\,\mu$m emission that is seen mainly towards the  northwest
of the cluster center, results most likely
from warm dust grains that are heated by the UV radiation of the 
B-type stars in the cluster. The $8\,\mu$m emission
traces the surface of somewhat denser cloud structures; it is
probably dominated by fluorescent emission from polycyclic aromatic hydrocarbon (PAH) molecules.
In the southeastern part of the cluster, the \textit{Spitzer} and 
\textit{Herschel} maps show a prominent pillar-shaped cloud, 
which points directly towards the M1.5Iab supergiant V361~Car (the very bright
point source in the Spitzer image). This pillar is very similar to the
numerous cloud pillars found in other parts of the CNC
\citep[see][]{Smith10b,McLeod16}.

The infrared images show a tendency of stronger cloud emission 
around the stellar cluster,
and less emission in the central parts of the cluster.
Together with the apparently empty bubble-like feature
just east of the cluster center, this suggest that the radiation and winds of the
high-mass stars in NGC~3293 have cleared the central regions of the
original clouds.
This conclusion is supported by the measured 
reddening values of stars with known spectral type in the NGC~3293 region
listed in \citet{Dufton06}; these values correspond\footnote{In their
optical photometric study, \citet{Baume03} 
found that the extinction law for the stars in NGC~3293
is  consistent with the usual relation $A_V = 3.1 \, E(B-V)$.}
to typical extinction values of 
$A_V \sim 1$~mag for stars near the cluster center,
whereas stars at the periphery show extinctions up to a few magnitudes.

\section{Random matching simulation\label{random-matches.sec}}

\begin{figure*}
\sidecaption
\includegraphics[width=12cm]{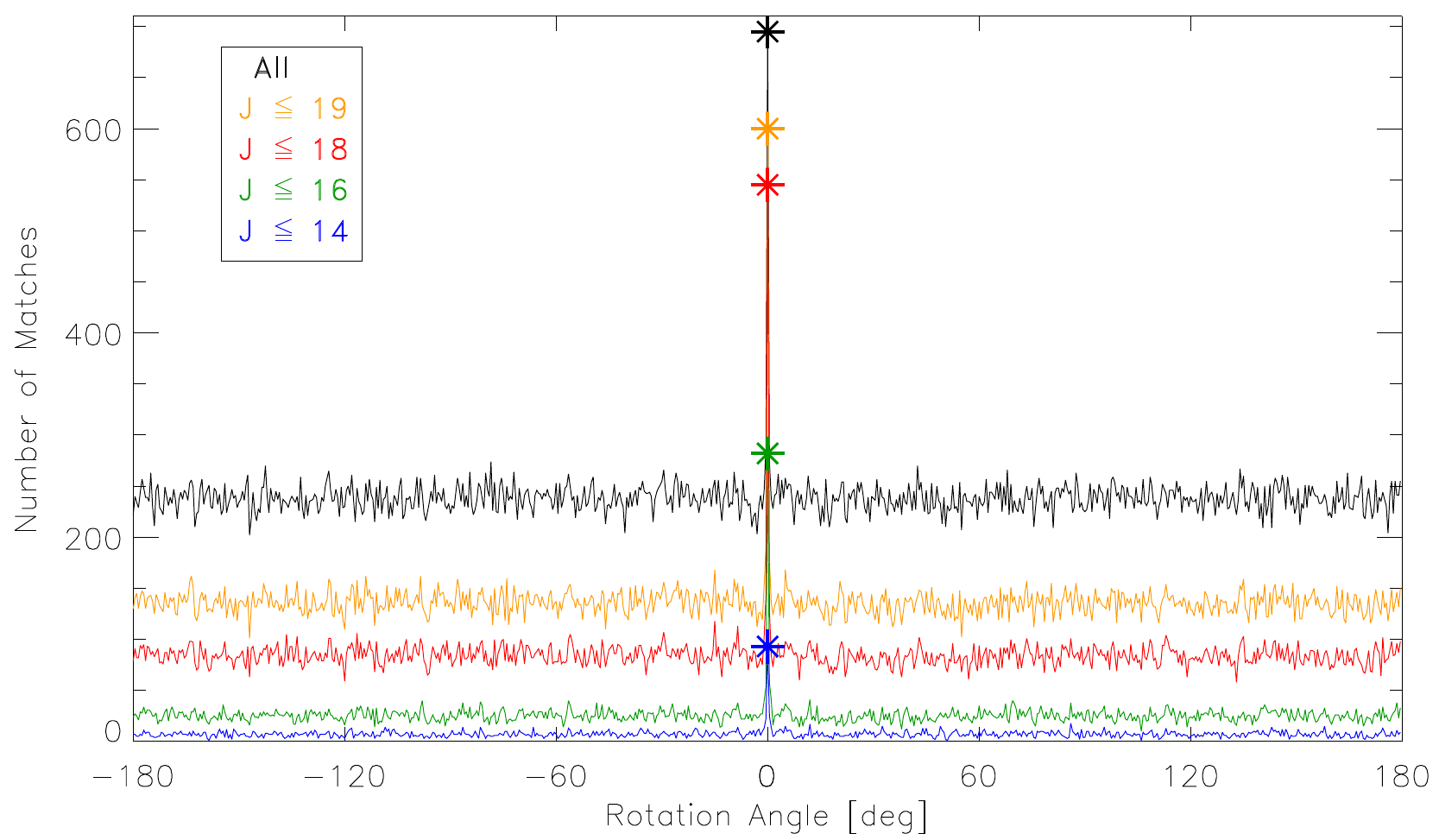}
\caption{Results of the random source matching simulations. The number of 
VISTA catalog matches is shown as a function of the rotation angle for
the X-ray source positions. The black line shows the results when using the
full VISTA catalog, and the colored lines show the results when using various restricted 
versions of the VISTA catalog with different magnitude limits. The thick asterisks mark the
number of matches found with the original (i.e.,~un-rotated) X-ray source catalog.
\label{matching-simulations.fig}}
\end{figure*}

In order to estimate the possible occurrence rate of false positive matches
between a random (spurious) X-ray source and an unrelated infrared source,
we performed a set of matching simulations.
We take into account that
neither the distribution
of the infrared sources, nor the distribution
and the properties of the X-ray matching regions are purely random, but
rather show characteristic spatial trends.

The distribution of the infrared sources in the field is not uniform
since we are looking at a stellar cluster with a higher density of relatively
bright stars in the center, superposed on an approximately uniform
distribution of relatively faint background objects,
and a reduced sensitivity for very faint objects in the
central regions around the very bright stars.

There are several reasons why the properties of the X-ray matching regions in a  \textit{Chandra}
observation are also not uniform over the observed field of view.
First, the sensitivity (and thus the detection limit)
 depends on the off-axis angle.
Second, the position uncertainties (and the corresponding radii
of the matching regions) also depend on the off-axis angle since
the point spread function increases with off-axis angle.
Third, the position uncertainties also depend on the number
of detected counts per source, which again depends on the 
local sensitivity (known as the effective area) at the source position
on the detector.
All the mentioned dependencies are, to first order, predominantly
functions of the angular distance from either the cluster center or the
focal point of the X-ray image; these two points coincide closely
in our observation.

We therefore use the original VISTA catalog for the  random matching
simulations, which preserves the spatial distributions of the
infrared sources. To simulate  X-ray source lists, we use our
original \textit{Chandra} ACIS-I source catalog in order to conserve
the angular-dependence of the source properties, but rotate all
source positions by certain angles around the
focal point in order to break the physical connection between X-ray
and infrared sources.
Simulations were performed with different rotation angles
from $-180\degr$ to $+180\degr$ in steps of $0.5\degr$.
At each rotation angle, we used {\tt match\_xy} in the same
way as for the original matching and recorded the number of
resulting matches with the VISTA catalog.
For any rotation angle that differs by more than a few degrees from
zero, the resulting match numbers are a measure of the expected number
of purely random
matches of sources without any physical connection.

The results of these simulations are shown in Fig.~\ref{matching-simulations.fig}.
For rotation angles of more than a few degrees,
it can be seen that the number of matches
shows only small fluctuations, and no significant variations
or trends with rotation angle,
confirming that the resulting numbers are good estimates of the
random false match fraction.

Considering the full VISTA catalog, the mean number of random
matches is $238 \pm 12$;
this corresponds to a
fraction of $(23.3 \pm 1.2)\%$ of all X-ray sources.

Two important aspects have to be considered in the interpretation 
of this number.
First, it should be noted that this number is a {strict upper limit}
to the actual number of possible random false matches in our sample 
since the simulation
assumes (by means of the rotation) that {no} X-ray source is 
physically connected to any infrared source; it is valid only for 
X-ray sources that  have no physical match detected in the available
optical/infrared images.
In reality, however, a large fraction of the X-ray sources are young stars,
i.e.,~rather bright IR sources, that produce correct positive matches.
Only those X-ray sources that have no true counterpart in the VISTA images
can get a false random match.

The second point concerns the magnitudes of these random false
matches.
The large majority of these are very faint infrared sources, most
of them with magnitudes $J > 18$.
This is just a reflection of the strongly increasing number
of infrared sources when going towards fainter magnitudes.
The color magnitude diagrams resulting from these
random matching experiments are therefore very different from
the actual color magnitude diagram, since they show only very few
random false matches to 
stars with sufficiently bright magnitudes to be 
considered as $\ga 1\,M_\odot$ stars.

To investigate this last point further, we repeated the simulations
with restricted versions of the original VISTA catalog, containing 
only objects above a specific magnitude limit.
The number of false matches drops steeply with increasing
magnitude limit:
for $J<18$ we find $84 \pm 9$ matches $(8.2 \pm 0.9\%)$,
for $J<17$ we find $47 \pm 7$ matches $(4.6 \pm 0.7\%)$,
and 
for $J<16$ we find $25 \pm 5$ matches $(2.3 \pm 0.5\%)$.
This implies that the fraction of potential false matches
of X-ray sources with stars that are bright enough to be considered 
$\ga 1\,M_\odot$ stars in NGC~3293 is quite small ($\la 4\%$).

\section{Random match with an artifact in the VST images\label{false-match.sec}}

\begin{figure}
\centering
\includegraphics[width=8.5cm]{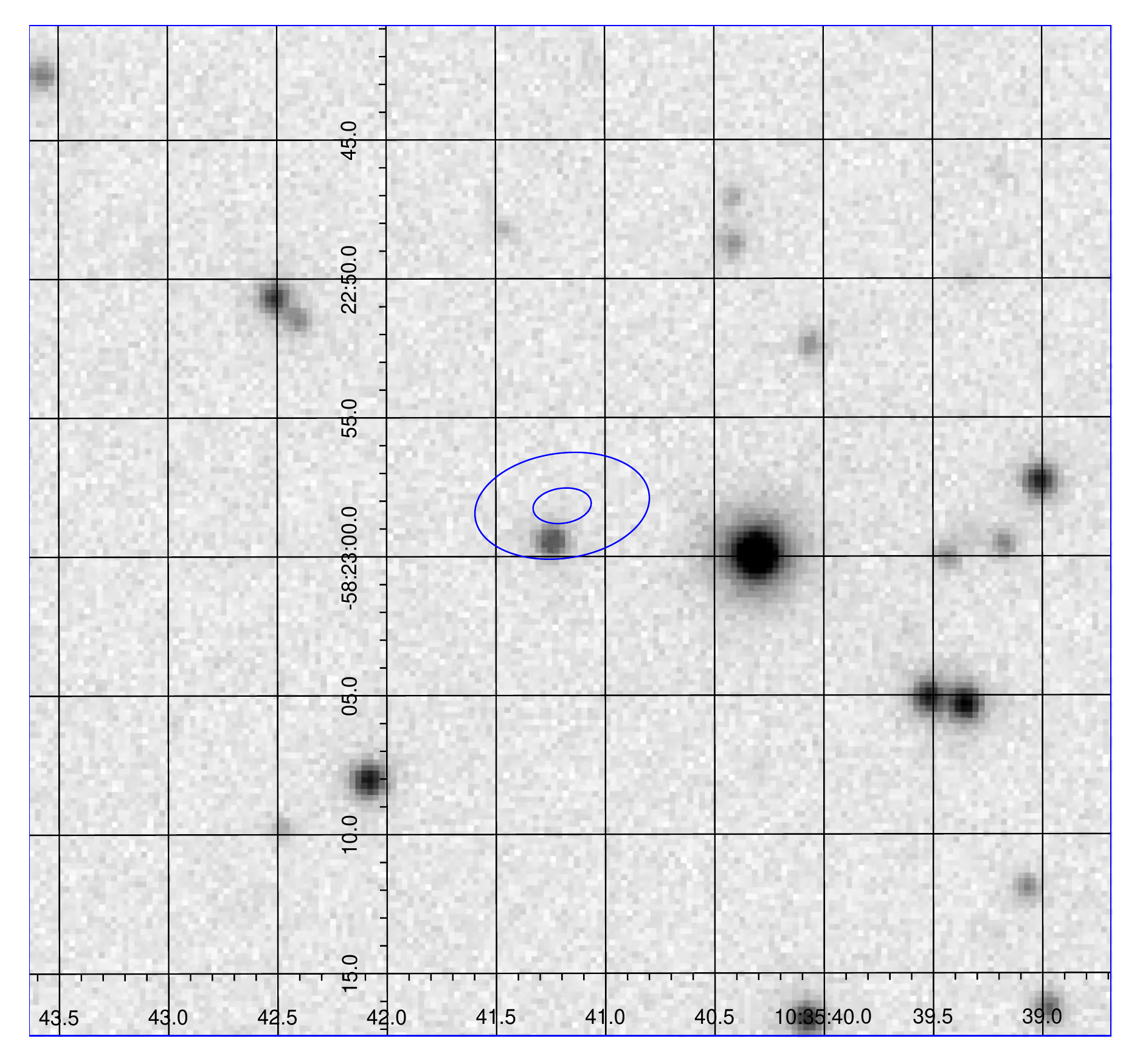}
\caption{VST $i$-band image of the region around the Chandra Source J103541.19-582258.1.
The small blue ellipse shows the $1\sigma$ error region for the X-ray source position;
the larger blue ellipse is  three times larger.
\label{artifact.fig}}
\end{figure}

A particularly interesting case of a random false match of an
X-ray source with an artifact in the VST optical catalog
concerns the 
Chandra Source 407, 
which is located 9 arcmin south of the cluster center.
The X-ray source yielded 12.8 net counts with a median photon energy
of $1.9$~keV.

The VST VPHAS-DR2 Point Source Catalogue 
lists a source 
VPHASDR2~J103541.2-582259.4 1735b-31-6789 with magnitudes
$r = 18.1483 \pm 0.026$, $ i = 17.9260 \pm 0.037$, 
and $H\alpha = 17.8223 \pm 0.022$, which
provides a possible match to the X-ray source.

The ESO archive contains 16 individual VST observations
of this point, obtained in the $u, g, r, i$, and the H$\alpha$ band.
Our inspection of all these
individual images yielded a very surprising result.
While we could clearly
confirm the presence of a point-like source in the $u$-band and $g$-band images
obtained on 15 Feb 2012 and in the $r$-band, $i$-band, and H$\alpha$ images obtained on
30 April 2012 (see Fig.~\ref{artifact.fig}),
all  other available VST images, which were
obtained on 15 Feb 2012 and 14 Feb 2013, show 
{no} source at this position.
The inspection of our VISTA images (obtained on 6 March 2012 and 8 March 2012)
and archival ESO WFI images obtained on 9 Feb 2012 also showed no source
at this position.
Upper limits for the magnitudes in these non-detection images are
$u> 21.8$,  $g > 22.5$, $r > 21.8$,  $i > 20.8$, and H$\alpha > 20.6 $.

These numbers  imply that the object  brightened by several magnitudes
in the six days from 9 Feb 2012 to 15 Feb 2012,
was invisible six days later, again bright one month later,
and again invisible in the most recent images. 
A possible explanation of such a very unusual behavior 
could be that the source is some kind of a transient object.

Finally, however, our
close inspection of those VST images in which the source
was visible, showed that the 2D profile of the object deviates 
from the normal point spread function of other sources in the surrounding.
After detailed consultations with VST instrument and imaging experts at the ESO headquarters,
it was finally found that the apparently highly variable object
is actually an image artifact, which was caused by an 
electronic  cross-talk effect
induced by a very bright, saturated star on another CCD chip in the camera
(private communication from the ESO User Support Astronomer M.~Petr-Gotzens 
and the ESO Optical detector engineering group).

Although it is very unlikely to find an optical artifact in the very
small matching region around an X-ray source, this case demonstrates that
these rare  coincidences can sometimes happen.
This highlights the importance of a careful
visual inspection of the original optical/infrared images
for a reliable determination of the counterparts to the X-ray sources.

\section{ Quasar candidate J103621.39-581520.0 \label{agn.sec}}

\begin{figure}
\centering
\includegraphics[width=8.5cm]{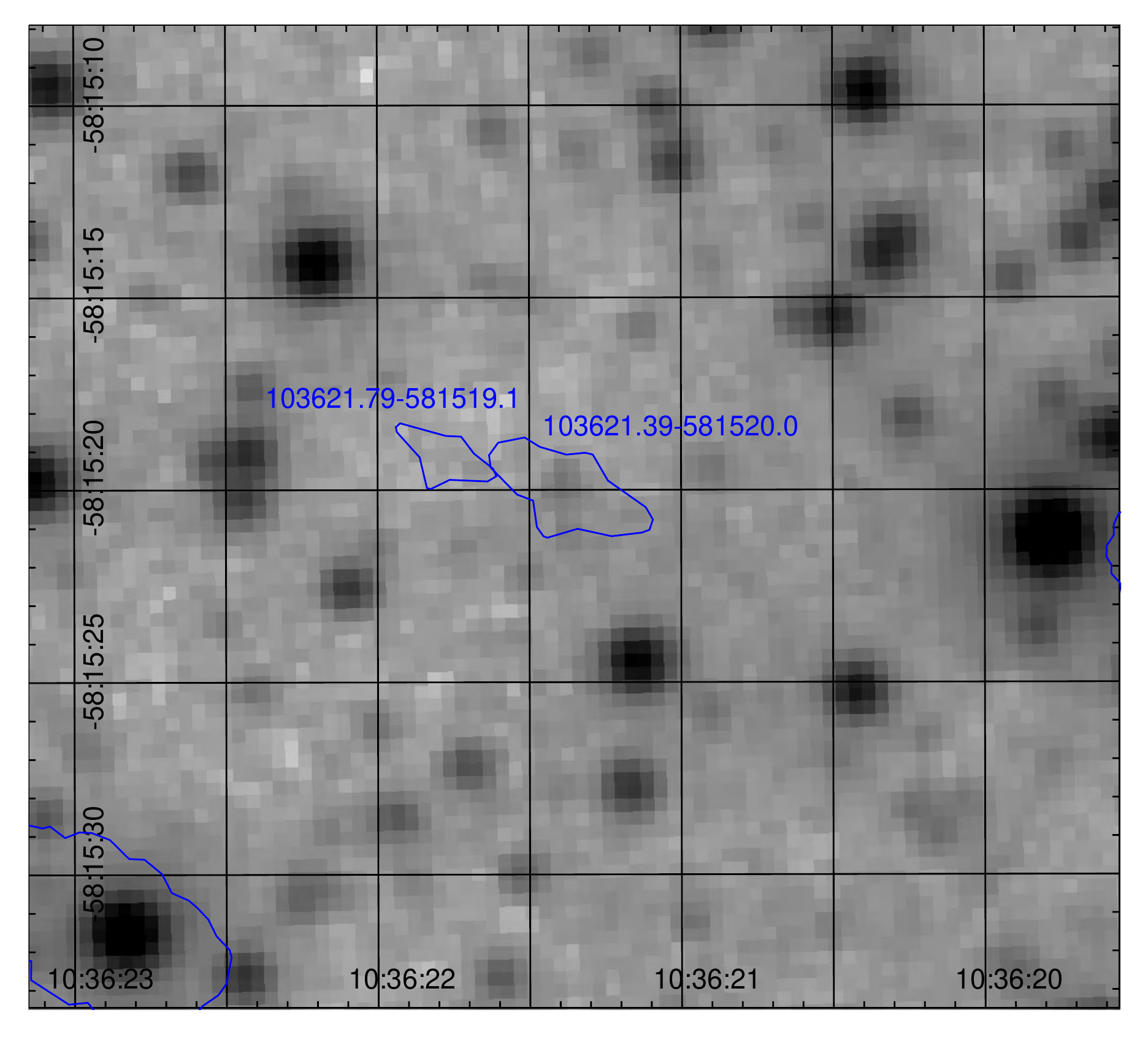}
\caption{VISTA $H$-band image of the region around the quasar candidate J103621.39-581520.0.
A logarithmic intensity scale was used to make the extremely faint infrared
counterpart of the X-ray source J103621.39-581520.0 visible.
The blue polygons show the X-ray extraction regions for the source.
\label{quasar-ima.fig}}
\end{figure}

As mentioned in Sect.~\ref{spectra.sec}, the brightest X-ray source (number 943)
in the ACIS-I array, with 190 net counts,
has no optical counterpart.
It is located $\approx 4.4\arcmin$ southeast of the cluster NGC~3293.
In Fig.~\ref{quasar-ima.fig} the \textit{Chandra} extraction region
of 103621.39-581520.0 is shown on the VISTA $H$-band image,
which reveals, at close inspection, an extremely faint infrared counterpart 
of this X-ray source.

The X-ray spectrum can be very nicely reproduced by a model
assuming a power-law X-ray spectrum and interstellar extinction.
The spectral fit yields  $\chi^2_r = 0.37$ for the following parameters:
$N_{\rm H} = (2.86 \pm 0.77) \times 10^{22}\,{\rm cm}^{-2}$
(corresponding to a visual extinction of $A_V \approx 14$~mag), 
and a 
photo index of $ 1.57 \pm 0.37$.
The observed $[0.5\!-\!8]$~keV X-ray flux, derived from the fit results, is 
$8.7 \times 10^{-14}\,{\rm erg}\,{\rm cm}^{-2}\,{\rm s}^{-1}$, and
the corresponding de-reddened X-ray flux
 is $1.5 \times 10^{-13}\,{\rm erg}\,{\rm cm}^{-2}\,{\rm s}^{-1}$.

The source is invisible in all available optical VST images.
From the VST $g$-band image, we estimate an upper limit of
$g \ge 23$~mag for the brightness of the source.
This corresponds to a flux of 
$ \le 4 \times 10^{-15}\,{\rm erg}\,{\rm cm}^{-2}\,{\rm s}^{-1}$.
Comparing this to the X-ray flux derived from the spectral fit
yields an X-ray to optical flux ratio 
$F_{\rm X} / F_{\rm opt} \ge 21$.

These properties, i.e., the combination of strong absorption,
a hard spectrum, and the very high X-ray-to-optical flux ratio,
are best explained by assuming the source to be an obscured quasar.
Many of these objects have X-ray-to-optical flux ratios above 1,
and a small fraction of about 5\% have $F_{\rm X} / F_{\rm opt} \ge 20$ \citep{DellaCeca15}.
J103621.39-581520.0 thus seems to be a particularly X-ray active quasar.

\begin{figure} \centering
\includegraphics[width=9.2cm,keepaspectratio]{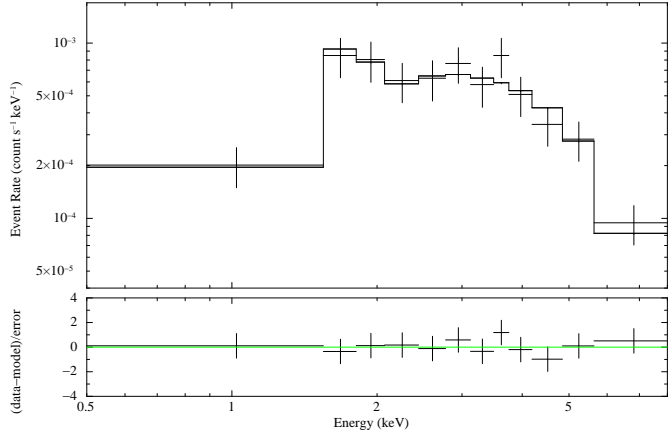}
\caption{X-ray spectral fit for J103621.39-581520.0 with a power-law model.
\label{quasars-spec.fig}}
\end{figure}

\end{appendix}

\end{document}